\documentclass[twocolumn]{aastex62}

\lefthead{Wada} 
\righthead{Obscuring fraction in AGNs}

\newcommand{\bea}{\begin{eqnarray} }
\newcommand{\eea}{\end{eqnarray}}

\begin{document}

\title{Obscuring Fraction of Active Galactic Nuclei: Implications from Radiation-driven Fountain Models}

\author{%
Keiichi Wada
}
\affiliation{Kagoshima University, Graduate School of Science and Engineering, Kagoshima 890-0065, Japan}
\affiliation{Ehime University, Research Center for Space and Cosmic Evolution, Matsuyama 790-8577, Japan}
\affiliation{Hokkaido University, Faculty of Science, Sapporo 060-0810, Japan}
\correspondingauthor{Keiichi Wada}
\email{wada@astrophysics.jp}

%




\begin{abstract}
Active galactic nuclei (AGNs) are believed to be obscured by an
optical thick ``torus'' that covers a large fraction of solid
angles for the nuclei. This torus paradigm has been widely accepted, but the physical
origin of the tori and the differences in the tori among AGNs are not clear.  In
a previous paper based on three-dimensional radiation-hydorodynamic
calculations, we proposed a physics-based mechanism for the
obscuration, called ``radiation-driven fountains,'' in which the
circulation of the gas driven by central radiation naturally forms
a thick disk structure that partially obscures the nuclear emission.  Here, we
expand this mechanism and conduct a series of simulations to explore
how obscuration due to pure radiative feedback 
depends on the properties of AGNs. We found that the
obscuring fraction $f_{obs}$ for a given column density toward the
AGNs changes depending on both the AGN luminosity and the black hole mass. In particular, $f_{obs}$ for
$N_{\rm H} \geq 10^{22}$ cm$^{-2}$ increases from $\sim 0.2$ to $\sim 0.6$ as a function of the X-ray luminosity $L_{\rm X}$
 in the $L_{\rm X} = 10^{42-44}$ ergs
s$^{-1}$ range, but $f_{obs}$ becomes small ($\sim 0.4$) above a luminosity 
($\sim 10^{45}$ ergs s$^{-1}$).  
The behaviorsof $f_{obs}$ can
be understood by a simple analytic model and provide insight into 
the redshift evolution of the obscuration. 
The simulations also show that for a given $L_{AGN}$, $f_{obs}$ is always
smaller ($\sim 0.2-0.3$) for a larger column density ($N_{\rm H} \geq 10^{23}$
cm$^{-2}$), which is a natural consequence that the dense gas is mostly
concentrated in a thin disk.  We also found that more than 70\% of
the solid angles can be covered by the fountain flows
when the Eddington ratio is high ($\gtrsim 0.3$) and the central accretion disk is inclined in
terms of the disk plane. This suggests a possible structure of the deeply obscured AGNs.

\end{abstract}


\section{INTRODUCTION}


Active galactic nuclei (AGNs) such as quasars and Seyfert galaxies 
are important objects in the cosmic history. 
The enormous radiation energy of AGNs
significantly influences the formation 
and evolution of galaxies and their environment.
The evolution of AGNs in high redshift has been studied 
by wide-field optical and X-ray surveys, and these studies have recently become 
statistically more reliable based on large samples \citep[e.g.,][]{ueda2014,paris2014,buchner2015, aird2015}.
However, because AGNs are believed to be surrounded 
by an optically thick material for a fraction of 
the solid angles, we do not necessarily observe all the radiation from the nuclei.
Because this obscuration may affect the apparent evolution of the luminosity function \citep{merloni2013},
the fraction of absorbed objects 
is an essential quantity not only for interpreting observations 
\citep[e.g.,][]{lawrenceelvis1982,lafranca2005,ballantyne2006,hasinger2008, han2012}
but also for theoretical models of AGNs evolution 
\citep[e.g.,][]{enoki2003,fanidakis2011, enoki2014}.

The fraction of obscured objects, hereafter called the obscured fraction $f_{obs}$,
can be estimated by various observational methods.
The nature of the obscuration should be different depending on 
the wavelength, central luminosity, and redshift,
suggesting a complicated nature and evolution of the environment around AGNs \citep[e.g.,][]{lawrence2010,merloni2014}. In particular, $f_{obs}$ inferred 
from optical and infrared observations is relatively high ($f_{obs} \sim 0.4-0.6$) \citep{roseboom2013,lusso2013,shao2013} and shows weak dependence on the luminosity, whereas X-ray samples suggest that $f_{obs}$ decreases with increasing X-ray luminosity \citep{ueda2003,hasinger2008,han2012,merloni2014, ueda2014}.
This case may be understood using the receding torus model \citep{lawrence1991,simpson2005}, but recent surveys with large samples ({\em XMM-Newton} and {\em Swift-BAT}) suggested more complicated behavior in terms of $L_{\rm X}$: 
$f_{obs}$ is small for both low and high luminosities (i.e., there is a peak for
a certain luminosity). 
The redshift evolution of $f_{obs}$ is also suggested by different groups,
although the results are not fully consistent each other \citep{burlon2011, brightman2011, ueda2014, buchner2015, aird2015}. 

%


The obscuring fraction as a function of the AGN luminosity and observed wavelength
should be related to the morphology, internal structures, and size of the torus.
If the obscured fraction of AGNs are related to the luminosity of AGNs, 
it would be natural to assume that such obscuring properties are 
directly caused by the radiative and mechanical feedback from the AGNs.
As discussed by \citet{elvis2012}, 
the weak correlation between the X-ray and optical obscuration 
might suggest that obscuration is occurred on different scales.
The optical property of the X-ray-selected AGNs suggest that
the radiation environment around the nucleus may determine the properties of the obscuration
\citep{merloni2014}.

The obscuration is not necessarily caused by static structures of the 
interstellar medium (ISM), such as the ``donut-like'' torus \citep{urry95}.
Recently, \citet{wada2012} showed that
radiation from the accretion disk drives vertical circulation of 
gas flows on the order of a few to tens of parsecs,
forming thick torus-like structures.
Based on this model, type-1 and -2 SEDs can be also consistently explained 
as a result of the outflows and backflows \citep{schartmann2014}.
This ``radiation-driven fountain'' has an advantage that
it naturally explains a long-standing question about the torus paradigm: how does the torus maintain its
thickness?  At a much smaller scale, the outflows originating in the accretion disk could obscure the nucleus \citep{elvis2000, proga2000,nomura2013}. Moreover,
\citet{elitzur06} also claimed
that the doughnut-like torus is not necessary, based on a similar picture.


In this paper, using three-dimensional 
radiation-hydrodynamic simulations, which are based on \citet{wada2012} and \citet{schartmann2014}, 
we investigate how the obscured fraction depends on the properties of AGNs, i.e.,
luminosity and black hole mass (or Eddington ratio).
In \S 2, the numerical methods and models are described, and the numerical results are shown in 
\S 3. The results are analytically discussed using a simple toy model in \S 4.
The origin of the obscuring fraction suggested by recent surveys 
are discussed in \S 5, and the summary and remarks for the present study are given in \S 6.
Here, we focus on the effect of the radiative feedback from the AGNs alone; 
possible synergetic effects with nuclear starbursts \citep[e.g.,][]{wada2002} 
will be discussed elsewhere.

\section{NUMERICAL METHODS AND MODELS}
\subsection{Numerical methods}
We follow the same numerical methods in this study as those presented in Paper I:
three-dimensional hydrodynamic equations are solved by an Eulerian hydrodynamic code with
a uniform grid, accounting for radiative feedback processes from the central source using a ray-tracing method.
The methods are briefly summarized here.

Based on our three-dimensional, multi-phase hydrodynamic model \citep{wada09}, 
we included the radiative feedback
from the central source, i.e., the radiation pressure on the surrounding gas 
and the X-ray heating. 
{In contrast to previous analytical and numerical studies of radiation-dominated tori \citep[e.g.,][]{pier93, shi-krolik2008,dorodnitsyn2011}, we assume neither dynamical equilibrium nor geometrical symmetry.
Outflows are naturally expected under the AGN feedback 
as also explored by \citet{dorodnitsyn2012}, in which infrared dominated flows in a pc-scale torus, 
by using time-dependent, axisymmetric radiation hydrodynamic simulations
 with a flux-limited diffusion.}

We further account for 
the self-gravity of the gas, radiative cooling, uniform ultraviolet (UV) radiation 
for photoelectric heating, and H$_{2}$ formation/destruction. However, to clarify only the effects of the radiation feedback, we do not include the effect of supernova feedback in this study. 

The hydrodynamic part of the basic equations is solved using 
the advection upstream splitting method (AUSM) \citep{liou93} and its 
improved scheme AUSM$^{+}$ \citep{liou1996}.
We use $128^3$ grid points for the hydrodynamics calculations.
The uniform Cartesian grid covers a $32^3$ pc$^3$ region
around the galactic center with a spatial resolution of 0.25 pc.
The Poisson equation is solved to calculate the
self-gravity of the gas using the fast Fourier transform, and the
convolution method uses $256^{3}$ grid points.



{The radiation source is an accretion disk whose size is 
five orders of magnitude smaller than the grid size in the present calculations.
Therefore, we assumed that the radiation is emitted from a point source.
However, the radiation flux originating from the accretion disk 
is assumed to be non-spherical (see \S 2.2). 
The optical depth $\tau_{\nu}$ is calculated at every time step along a ray from the central source at
each grid point; $128^{3}$ rays are used in the computational box.
We assume a cooling function based on a
radiative transfer model of photodissociation regions
\citep{meijerink05} for 20 K $< T_{gas} < 10^{8}$ K, 
which is a function
of the molecular gas fraction $f_{\rm H2}$ and the far ultraviolet (FUV) intensity $G_0$ \citep{wada09}.
Here, we assume solar metallicity and $G_{0} = 1000$.
The main radiative heating source for the ISM is the X-ray, which reflect
interactions between high-energy secondary electrons
and thermal electrons (Coulomb heating) as well as H$_{2}$ ionizing heating
in the warm and cold gases \citep{maloney96, meijerink05}.

The temperature of the interstellar dust $T_{\rm dust}$ at a given position irradiated by the central source
is calculated by assuming thermal equilibrium \citep[e.g.,][]{nenkova2008}. 
If $T_{\rm dust} > 1500\,{\rm K}$, we assume that the dust is sublimate.
Here, we assume that the ISM in the central several tens parsecs is optically thin 
with respect to the re-emission from the host dust.
The SED of the AGN and the dust absorption cross-section are taken from \citet{laor1993}.
We use a standard galactic dust model, which is the same as that used by \citet{schartmann2010}.

\subsection{Initial conditions and model parameters}
We follow the three-dimensional evolution of a rotating gas disk in a
fixed spherical gravitational potential under the influence of radiation from 
the central source.
To prepare quasi-steady initial conditions without
radiative feedback, we first run models of an axisymmetric and rotationally supported thin
disks with uniform density profile.
After the thin disks are dynamically settled, the radiation 
feedback is turned on.

%
%


Free parameters in the present simulations is the luminosity of the AGN ($L_{AGN}$) 
and the black hole mass ($M_{\rm BH}$).
The Eddington ratios $L_{AGN}/L_{E}$, 
for the Eddington luminosity $L_{E} \equiv  4\pi G c m_{p} M_{\rm BH}/\sigma_{T} $,
are changed from 0.01 to 0.5 for three black hole masses:  
$M_{BH,8}\equiv M_{BH}/10^{8}M_{\odot} = 0.13, 1.3$, and 13. 
The total AGN luminosity ranges $ 1.6\times10^{43} - 8.0 \times 10^{46}$ erg s$^{-1}$,
which correspond to X-ray luminosities of $L_{X, 44} \equiv   L_{\rm X}/(10^{44}\; {\rm ergs}\; {\rm s}^{-1} )
= 0.01-8.35$. 
We also prepare the initial gas disk with two different densities: $\rho_{0} =1.2$ and $3.6 \times 10^{3} M_{\odot}\; {\rm pc}^{-3}$.  
Thus, the gas mass fraction to the black hole mass ranges from $4 \times 10^{-4}$ to 0.12.
The AGN luminosity is assumed to be constant during calculations, which is typically a few million years.

{
We have modified some assumptions in Paper I to achieve more realistic models.
As is in Paper I, the accretion disk itself is not solved in present simulations,
and we assume an anisotropic radiation field for the 
central source. 
In contrast to the assumption in Paper I, i.e. $F_{AGN} \propto \cos \theta$ for all wavelength,
here the ultraviolet flux changes as $F_{UV}(\theta) \propto \cos \theta (1+2\cos \theta)$ \citep{netzer1987},
where $\theta$ denotes the angle from the rotational axis ($z$-axis)
in order to take into account the limb darkening effect of the thin accretion disk. 
As a result, the UV flux is about 20\% larger for the $z$-axis and 25\% smaller at $\theta \sim 60^\circ$
than the case in Paper I. Since the X-ray radiation would be originated in the hot coronal atmosphere of
the accretion disk \citep{netzer1987, xu2015}, here we simply take 
spherical symmetry for the X-ray flux.
In Paper I, $L_X$ was fixed to
10\% of the bolometric luminosity, 
but here we use luminosity-dependent correction for the X-rays \citep{hopkins2007} and for the $B$-band \citep{marconi2004}.
The X-ray luminosity for $M_{BH}=1.3\times 10^7 M_\odot$ is about half of that assumed in paper I.}
{In paper I, we also simply assumed that the hypothetical accretion disk is parallel to the
galactic disk. However, in reality, any orientations of the accretion disk could be possible 
in dependent of the galactic disk \citep[e.g.][]{kawaguchi2010}.
Here we assumed that the direction of the emerging non-spherical radiation
is inclined by 45$^\circ$ from the rotational axis of the circumnuclear gas disk in some models, and 
its three-dimensional gas dynamics is fully followed without assuming any symmetry.
} 

{The dust-to-gas ratio is assumed to be a widely used value, i.e. 0.01, instead of 1/30 in Paper I,
although the actual ratio is not clear in the AGN environment. 
}

As summarized in Table 1, we explored a total of 24 different sets of parameters (each model name 
represent parameters as described in the caption).

%
%
%

\section{NUMERICAL RESULTS: Obscuration by Radiation-Driven Fountain}

\begin{figure*}[h]
\begin{center}
\includegraphics[width = 12cm]{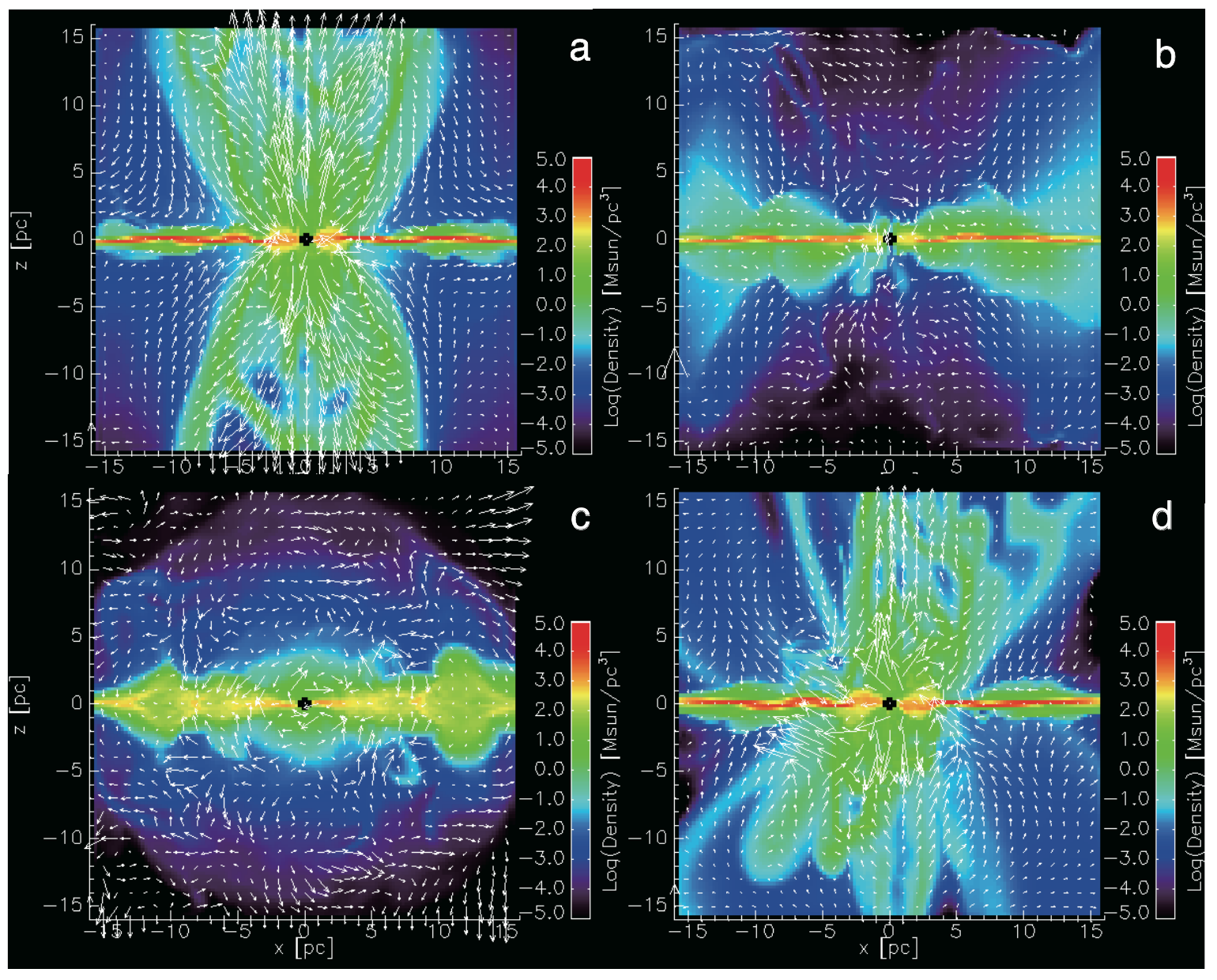} \\
\caption{(a) Density distribution of the x-z plane for model 8D50 (see Table 1)
with the X-ray luminosity $L_{X,44} = 1.51$, black hole mass $M_{BH,8} = 1.3$, Eddington ratio ($\gamma_{Edd}$) = 0.5. The central accretion disk inclination $\phi_{i} =0$. The bipolar outflows and back flows toward the disk (``radiation-driven fountain'') 
are formed.
(b) Same as (a), but for model 7L01 with $(L_{X,44}, M_{BH,8}, \gamma_{Edd}) = (0.01, 0.13, 0.01)$. No clear outflows are formed.
(c) Same as (a), but for model 9L40 with $(L_{X,44}, M_{BH,8}, \gamma_{Edd}) = (7.08, 13.0, 0.4)$.
(d) Same as (a), but for model 8D30i with $(L_{X,44}, M_{BH,8}, \gamma_{Edd}) = (1.03, 1.3, 0.3)$, 
and the central accretion disk is inclined by 45$^{\circ}$ from the $z$-axis ($\phi_{i} = \pi/4)$).
The arrows represent the relative velocity of the gas on the same plane. (Note that the unit size of the vectors is
different for each figure.)}
\end{center}
\label{wada_fig: f001}
\end{figure*}

\begin{figure*}[h]
\centering
\includegraphics[width = 8cm]{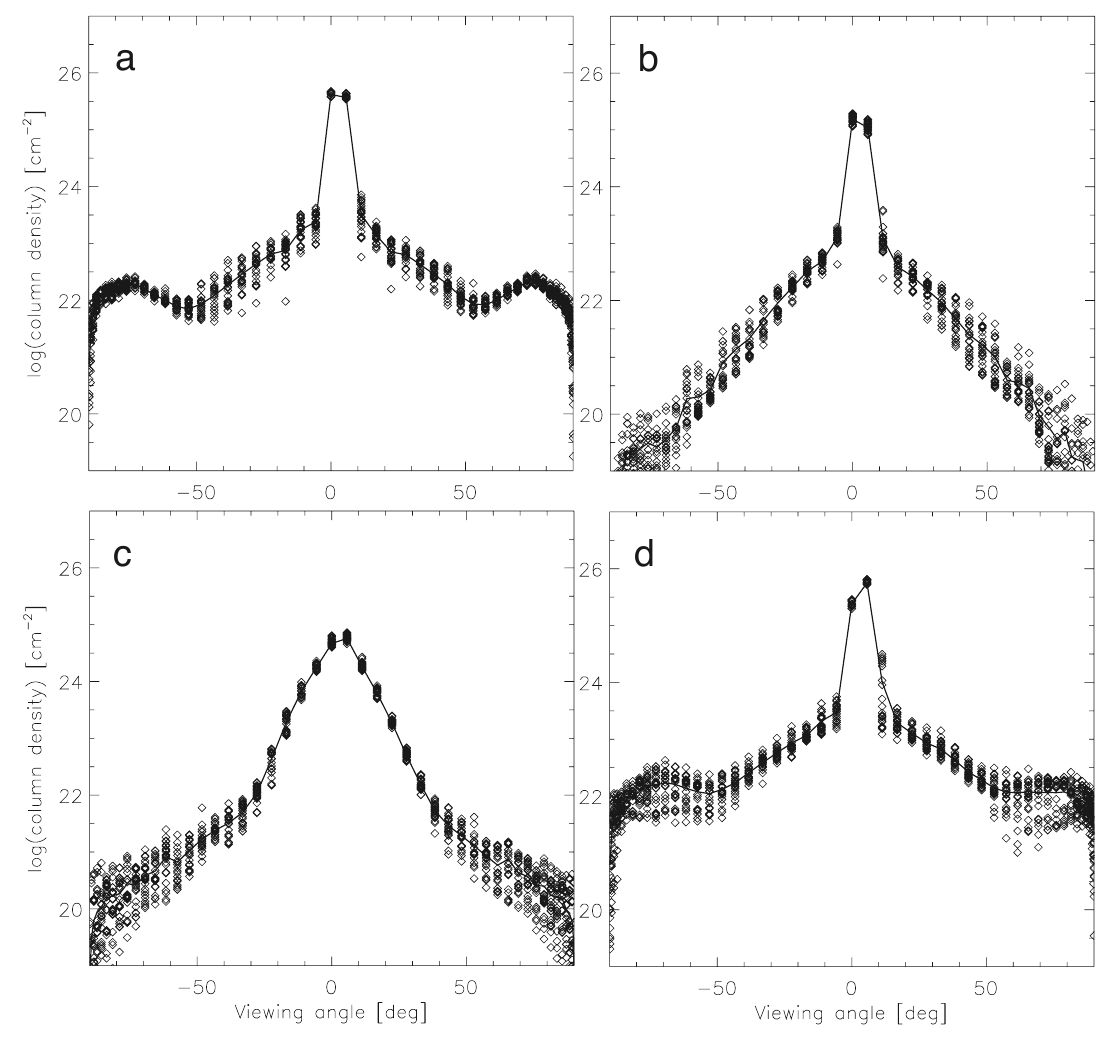} \\
\caption{Column density distributions toward the nucleus as a function of 
the viewing angle ($0^{\circ}$: edge-on, $\pm 90^{\circ}$: pole-on) for the same four models shown in Fig. 1 (a: 8D50, b: 7L01, c: 9L40, and d:8D30i). For a given viewing angle, the dispersion of $N_{\rm H}$
reflects the inhomogeneous density structure of the circumnuclear material. 
The solid line is the azimuthal average.}
\label{wada_fig: f002}
\end{figure*}

\begin{figure}[h]
\centering
\includegraphics[width = 8cm]{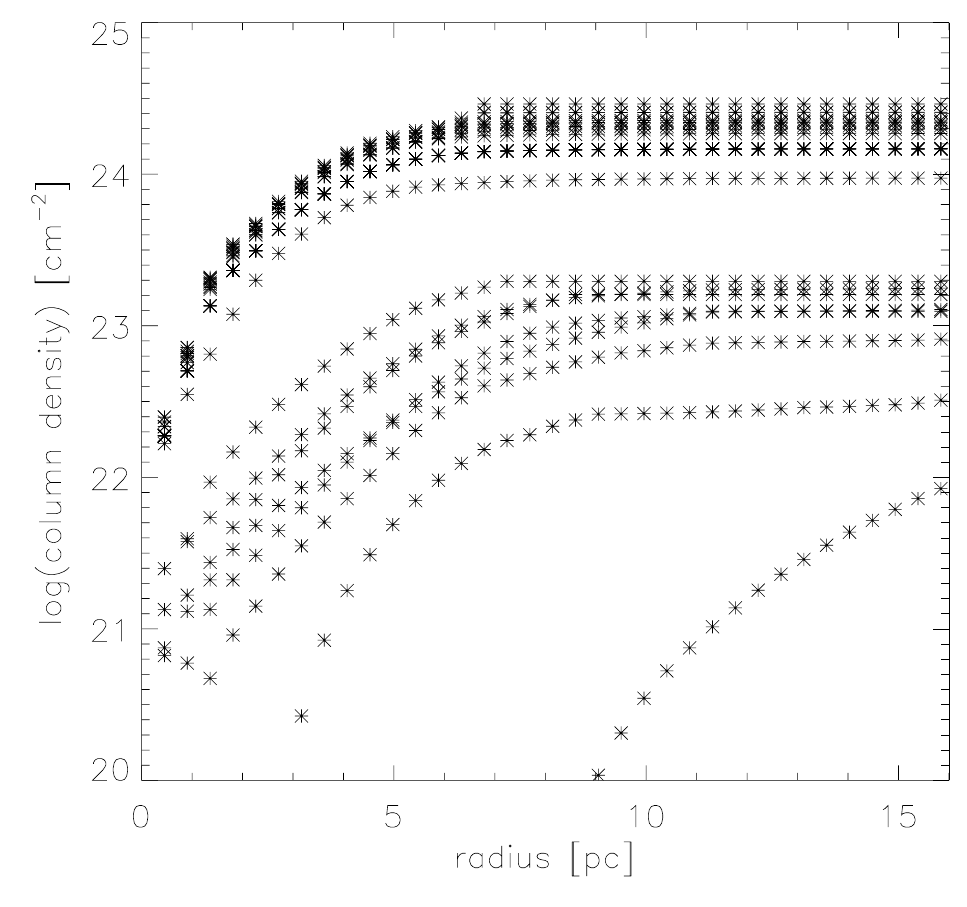} \\
\caption{Radial distribution of cumulative column densities in model 8D50 (Fig.1a and Fig.2a). The column densities
are evaluated along line-of-sights toward the nucleus every about 7 degrees between the rotational axis and the disk plane. 
It shows that most of the obscuration occurs by the gas inside $r \sim  10$ pc.}
\label{wada_fig: colden_dist}
\end{figure}

\begin{figure}[h]
\begin{center}
\includegraphics[width = 8cm]{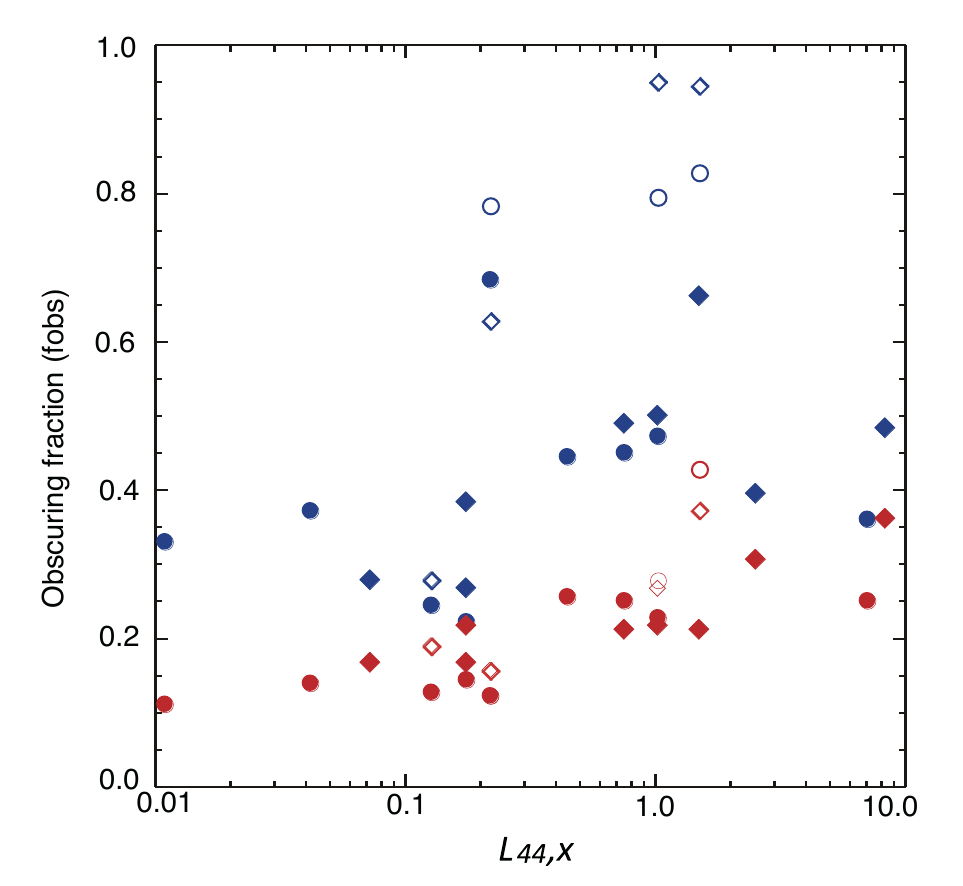} 
\caption{Obscuration fraction as a function ($f_{obs}$) of the X-ray luminosity in 24 models.
In each model, $f_{obs}$ is plotted for  $N_{22} \equiv 10^{22}\; {\rm cm}^{-2}$ (blue symbols) and for $N_{23} = 10^{23}\; {\rm cm}^{-2}$ (red symbols). 
Circles represent initial gas disks with $\rho_{0} = 0.4\times 10^{3} M_{\odot}\; {\rm pc}^{-3}$, and diamonds represent those with $\rho_{0} = 1.22\times 10^{3} M_{\odot}\; {\rm pc}^{-3}$.
Open symbols represent $\phi_{i} = \pi/4$; that is, the central accretion disk is inclined by 45$^\circ$ (i.e. 
model names with ``i''. See also  Fig. 1d and 2d).}
\end{center}
\label{wada_fig: f007}
\end{figure}

In this section, we derive the obscuring fraction
for a given column density toward the nucleus using a quasi-steady three-dimensional density distribution
under the influence of the radiation feedback.
The results are summarized as a function of the X-ray luminosity.

Four representative models of the density structures of the gas
at a quasi-stable state are shown in Fig. 1.
Figure 1a shows a typical radiation-driven fountain of the gas (Paper I),
which consists of bipolar, non-steady, non-uniform outflows, failed winds (back flows to
the disk), and a dense, thin disk. 
{Physical origin of the fountain flows was partly discussed in paper I, but 
here we add some analysis of the numerical results in Appendix A-1.} See 
also the simple model in \S 4 for conditions to generate outflows. {As discussed in Appendix A-1.2, the system reaches to a quasi-steady
state of the gas circulation after about 2 rotational period at $r=10$ pc,
 which is $t\sim 0.5$ Myr for $M_{BH} = 1.3\times 10^3 M_\odot$ .}

Using the density distribution, the column density toward the nucleus
as a function of the viewing angles ($\theta_{v}$) and the azimuthal angles
is plotted in Fig. 2a. In Fig.2, column densities for the models shown in
Fig. 1 are also plotted.
Among these models,  
the column density tends to increase toward $\theta_v=0$ (edge-on) 
and reaches the maximum of $N_{\rm H} \geq 10^{25}\;{\rm cm}^{-2}$.
This occurs because of the stable existence of the dense, thin gas disk for which the radiation 
from the central source does not affect the disk except for the central part.
 The bipolar outflows and the failed winds cover a fraction of solid angles for the central source
 in model 8D50 (Fig.1a). 
Therefore, the nucleus is obscured
for $|\theta_{v}| \lesssim 90^{\circ}$ 
and  $|\theta_{v}| \lesssim 20^{\circ}$ for $N_{22} \equiv 10^{22} \; {\rm cm}^{-2}$
 and $N_{23} = 10 N_{22}$, respectively.
The density change for the azimuthal direction causes 0.5--1.0 dex difference in 
the column density for a given viewing angle, which causes a small ambiguity
for the obscured fraction ($f_{obs}$) as a function of the column density.

{
Figure 3 shows radial distribution of the cumulative column densities in a typical 
fountain model (8D50), from which we can know where the obscuration mostly occurs for a given line-of-sight.
It shows that $N_H \gtrsim 10^{22}$ cm$^{-2}$ at $r=16$ pc (i.e. the edge of the computational box)
 is mostly attained the gas inside 5-10 pc,
whose structures are resolved by 20-40 numerical grid points.
}

In some low luminosity models (e.g., model 7L01: $L_{X,44} = 0.011$ and $M_{BH,8}=0.13$) no steady 
outflows are observed 
(Fig. 1b), and
 as a result, the opening angle of the obscuring material becomes large with $ f_{obs}(N_{22})  \equiv f_{obs,22} = 0.33$ (Fig. 2b). 
On the other hand, 
the obscuring fraction also becomes small ($f_{obs,22} \simeq 0.4$) for high luminosity with
a massive black hole (model 9L40: $L_{X,44} = 7.08$ and $M_{BH,8} = 13$), such as those 
shown in Fig. 1c and 2c. These high luminosity models 
do not also show steady bipolar outflows.
This is because the acceleration due to the radiation pressure for dusty gas is not sufficiently large 
to reach the escape velocity for the luminous models, in which 
the dust sublimation radius moves outward ($\propto L_{AGN}^{1/2}$).
This is discussed using a kinematic model in \S 4.

As a result of the fully three-dimensional calculations, the central radiation source (the accretion disk)
is not restricted to a parallel configuration with the rotational axis of the gas disk.
In reality, the accretion disk should not necessarily be in the same same plane of the circumnuclear gas disk.
We found that if the axis of the central source is inclined by 45$^\circ$ from the rotational axis of the disk, 
and the Eddington ratio is relatively high ($\gtrsim 0.3$), the majority of the solid angle for the AGN
is covered by the fountain gas (Fig. 1d and 2d).

The obscuring fraction of all 24 models are summarized in Table 1 and Fig. 4, where
$f_{obs}$ for $N_{22}$ and $N_{23}$ are plotted for each model
as a function of $L_{X,44}$. 
Figure 4 shows that nature of the obscuration depends on the luminosity of AGNs, 
and $f_{obs,22}$ is small in 
both low and high luminosity cases. 
Moreover, $f_{obs,22}$ shows a maximum value of 0.8--0.9 around $L_{X,44} \sim 0.2-1.0$.
The obscuring fraction for $N_{23}$ gradually increases from $f_{obs,23} \simeq 0.1$ to 0.3 in
a range of $L_{X,44}=0.01 - 10$, which is almost the same as the initial conditions.
 This reflects that the scale height of the dense part of the disk
is larger for more luminous models (i.e., more massive black holes).
For a less dense column density, $f_{obs,22}$ increases from 0.3 to 0.5--0.6 
toward $L_{X,44} \sim 1.0$,
as a consequence of the formation of radiation-driven fountains.
There is no significant difference on $f_{obs}$ between models with different initial gas density $\rho_{0}$.

We also find that $f_{obs,22}$ becomes quite large  ($\sim 0.7-0.9$) 
for parameters that cause outflows ($\gamma_{edd} \gtrsim 0.3$ and $M_{BH,8} =$ 0.13 or 1.3)
when the central accretion disk
is inclined by 45$^{\circ}$ (Fig. 1d, 2d, and open blue symbols in Fig. 4). 
This occurs because the gas circulation between the outflows and
the dense gas effectively covers a large fraction of the solid angle.
On the other hand, for more luminous AGNs ($L_{X,44} > 2$), the outflows
are stalled; therefore, a relatively small obscured fraction is observed.
For a given parameter set, note that $f_{obs}$  should have an ambiguity of $\sim 10$\%,
which comes from the azimuthal inhomogeneous structures of the fountains (see Fig. 2).


{Although the overall structures of $N_{\rm H}$ presented here is similar to
those found in paper I, there are some differences, because of the modified assumptions (\S 2.2).
The radiative feedback assumed in this paper is less effective for generating outflows
than those in paper I for the same BH mass.
Moreover the present models show smaller dispersion of $N_{\rm H}$ 
for a given viewing angle $\theta$, especially for smaller $\theta$.
This is partly because the circumnuclear disks in the present model keep its geometrically thin
shape in the computational box. On the other hand, the disks are followed by 
twice larger computational box in paper I (i.e. $64^3$ pc). 
The disks tend to be gravitational unstable in the outer part (see discussion in Appendix A-1.2),
where gravity of the BH is weaker, resulting in geometrically thick, inhomogeneous disk.
This thick disks are not uniform for the azimuthal direction (see for example Fig. 7 in paper I), 
therefore the column density for a given viewing angle show dispersion. 
Although most obscuration is occurred in the central 10 pc, distributions and structures in 
the outer part of the disk on several tens parsecs scale may contribute to the observed column density, 
especially for the viewing angles close to the edge-on. 
Therefore, $N_{\rm H}$ could have larger dispersions, and as a result
the obscuring fraction especially for $N_{\rm } \gtrsim 10^{23}$ cm$^{-2}$ presented 
here might be a ``lower'' limit with the contribution of the attenuation by the gas 
on larger scales.
}

\begin{table*}[htbp]
\begin{center}
\caption{Model parameters and resultant obscured fraction.} 
\begin{tabular}{lllllllllll} \hline 
Models & $M_{BH}^{a}$ & $\gamma_{Edd}^{b}$ & $L_{AGN,46}^{c}$ & $L_{X,44}^{d}$ & $\phi_{i}^{e}$ & $\rho_0^{f}$ & $f_{obs,23}^{g}$ & $f_{obs,22}^{h}$  \\  \hline
$\ast$7L01 & 0.13 & 0.01 & 0.0016 & 0.01 & 0 & 0.40 & 0.11 & 0.33\\ 
7L05 & 0.13 & 0.05 & 0.008 & 0.04 & 0 & 0.40 & 0.14 & 0.37\\ 
7D30 & 0.13 & 0.3 & 0.016 & 0.18 & 0 & 1.22 & 0.17 & 0.28\\ 
7L20 & 0.13 & 0.2 & 0.032 & 0.13 & 0 & 0.40 & 0.13 & 0.24\\
7D20i & 0.13 & 0.2 & 0.032 & 0.13 & $\pi/4$ & 1.22 & 0.19 & 0.28\\
7L30 & 0.13 & 0.3 & 0.048 & 0.18 & 0 & 0.40 & 0.14 & 0.22\\ 
7D30 & 0.13 & 0.3 & 0.048 & 0.18 & 0 & 1.22 & 0.22 & 0.38\\ 
7D10 & 0.13 & 0.3 & 0.048 & 0.07 & 0 & 1.22 & 0.17 & 0.27\\ 
7L40 & 0.13 & 0.4 & 0.064 & 0.22 & 0 & 0.40 & 0.12 & 0.68\\ 
7D40i & 0.13 & 0.4 & 0.064 & 0.22 & $\pi/4$ & 1.22 & 0.16 & 0.63\\
7L40i & 0.13 & 0.4 & 0.064 & 0.22 & $\pi/4$ & 0.40 & 0.12 & 0.78\\ 
8L10 & 1.3 & 0.1 & 0.16 & 0.45 & 0 & 0.40 & 0.26 & 0.44\\
8L20 & 1.3 & 0.2 & 0.32 & 0.76 & 0 & 0.40 & 0.25 & 0.45\\
8D20 & 1.3 & 0.2 & 0.32 & 0.76 & 0 & 1.22 & 0.21 & 0.49\\
8L30 & 1.3 & 0.3 & 0.48 & 1.03 & 0 & 0.40 & 0.23 & 0.47\\
8D30 & 1.3 & 0.3 & 0.48 & 1.03 & 0 & 1.22 & 0.22 & 0.50\\
$\ast$8D30i & 1.3 & 0.3 & 0.48 & 1.03 & $\pi/4$ & 1.22 & 0.27 & 0.95\\ 
8L30i & 1.3 & 0.3 & 0.48 & 1.03 & $\pi/4$ & 0.40 & 0.28 & 0.79\\
8L40i & 1.3 & 0.4 & 0.64 & 1.51 & $\pi/4$ & 0.40 & 0.43 & 0.83\\
$\ast$8D50 & 1.3 & 0.5 & 0.8 & 1.51 & 0 & 1.22 & 0.21 & 0.66\\ 
8D50i & 1.3 & 0.5 & 0.8 & 1.51 & $\pi/4$ & 1.22 & 0.37 & 0.94\\
9D10 & 13 & 0.1 & 1.6 & 2.54 & 0 & 1.22 & 0.31 & 0.39\\
$\ast$9L40 & 13 & 0.4 & 6.4 & 7.08 & 0 & 0.40 & 0.25 & 0.36\\ 
9D50 & 13 & 0.5 & 8.0 & 8.35 & 0 & 1.22 & 0.36 & 0.48 &  & \\ \hline
\end{tabular} \end{center}
 $a$: Black hole mass ($10^{8}M_{\odot}$), $b$: Eddington ratio, $c$: Total AGN luminosity ($10^{46}$ erg s$^{-1}$),
 $d$: X-ray luminosity ($10^{44}$ erg s$^{-1}$), $e$: inclination angle of the accretion disk, $f$: Density of initial disk ($M_\odot \; {\rm pc}^{-3}$, $g$: Obscuring fraction for column density $N_{H}\leq 10^{23}$ cm$^{-2}$, $h$: Obscuring fraction for column density $N_{H}\leq 10^{22}$ cm$^{-2}$.  Each model name represents ``black hole mass'' (7, 8, or 9) $+$ density of the
 initial gas disk (L or D) $+$ Eddington ratio (0.05-0.5) $+$ inclination of the accretion disk ($i$: $\phi_{i}=\pi/4$). Models with $\ast$ in their names are shown in Fig. 1 and 2.

\end{table*}

\section{A SIMPLE ANALYTIC MODEL}

As shown in \S 3 and paper I, the three-dimensional gas dynamics under the effect of the 
non-spherical radiation are generally complicated.
However, a simple kinematic model is helpful for understanding 
the behavior of the radiative feedback and the origin of the dependence of $f_{obs}$ 
on $L_{AGN}$ in Fig. 4.

In the phenomena discussed here, the primary driving force producing the fountain of gas
around the AGNs is the radiation pressure for the dusty gas.
We assume that the accretion disk radiates most of its energy toward the 
direction of the rotational axis, not toward 
the plane of the gas disk (except for the case of the inclined accretion disk). 
 On the other hand, the radiative heating more isotropically affects the
 surrounding gas through advection.
The X-ray radiation from the AGN, which is assumed to be spherically symmetric
in this study, heats the inner part of the thin gas disk, 
making it geometrically thick.
Thus, we expect that the gas is radially accelerated by the radiation 
pressure with assistance from the X-ray heating.
In the following discussion, the effect of the gaseous pressure in the outflowing gas and 
the self-gravity of the gas are ignored for simplicity.
The outflowing gas is also dealt as a ``parcel''  or ``shell'', that is
the attenuation of the central radiation by the outflows themselves are ignored.

\begin{figure}[h]
\centering
\includegraphics[width = 5cm]{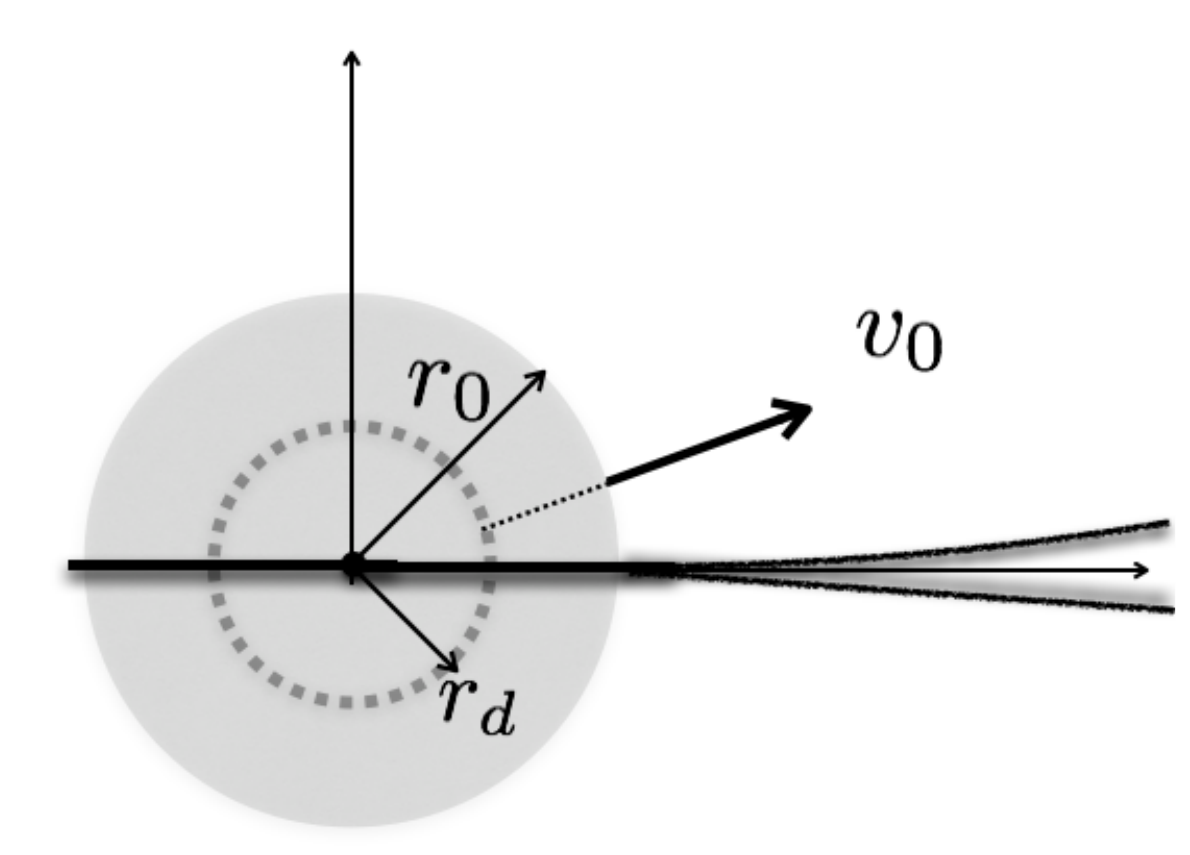} 
\caption{Schematic of the radiation-driven outflows, where $r_{d}$ is the dust sublimation radius, $r_{0}$ is
the thermal equilibrium radius in terms of the X-ray heating, and $v_{0}$ is the velocity of the dusty gas at $r=r_{0}$. See discussion in \S 4. }
\label{wada_fig: f008}
\end{figure}

\begin{figure}[h]
\centering
\includegraphics[width = 6.0cm]{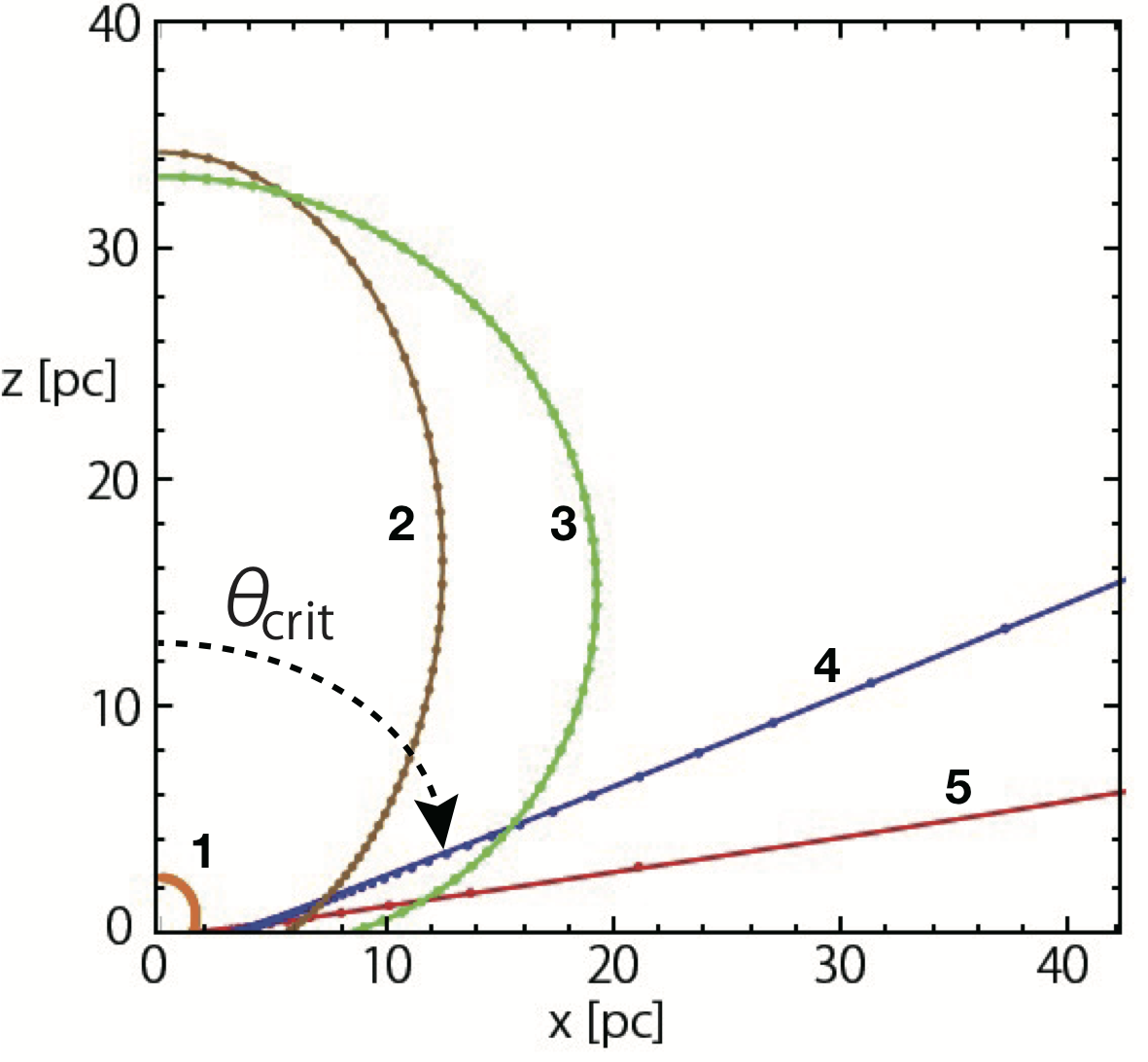} 
\caption{The maximum radius of the radiation-accelerated gas
$r_{max}$ ($L_{AGN}(\theta), M_{BH}$) (eq. \ref{eqn: 3-6}) on a $x-z$ plane for
five parameter sets: $(L_{46}, M_{BH,8})=$(0.02,0.1), (0.2,1.0), (10.0,10.0), (2.0,10.0), and (0.01,1.0), 
with red (denoted by no. 5), blue (4), green (3), brown (2), and orange (1) solid lines, respectively.
If $r_{max}$ are finite values for any angles (e.g., no. 1, 2, and 3), the escape velocity of 
the outflow gas is not large enough, then no steady outflows 
are expected. On the other hand, if $r_{max}$ becomes infinity for $\theta < \theta_{crit}$, 
as for no. 4 and 5, we expect bipolar outflows with a half-opening angle of $\theta_{crit}$,
and ``failed winds'' for launched gas with finite $r_{max}$.
For these cases, we suppose that 
the obscuring fraction $f_{obs}$ is close to $1- 2\theta_{crit}/\pi$.
}
\label{wada_fig: f1}
\end{figure}

\begin{figure}[h]
\centering
\includegraphics[width=8.0cm]{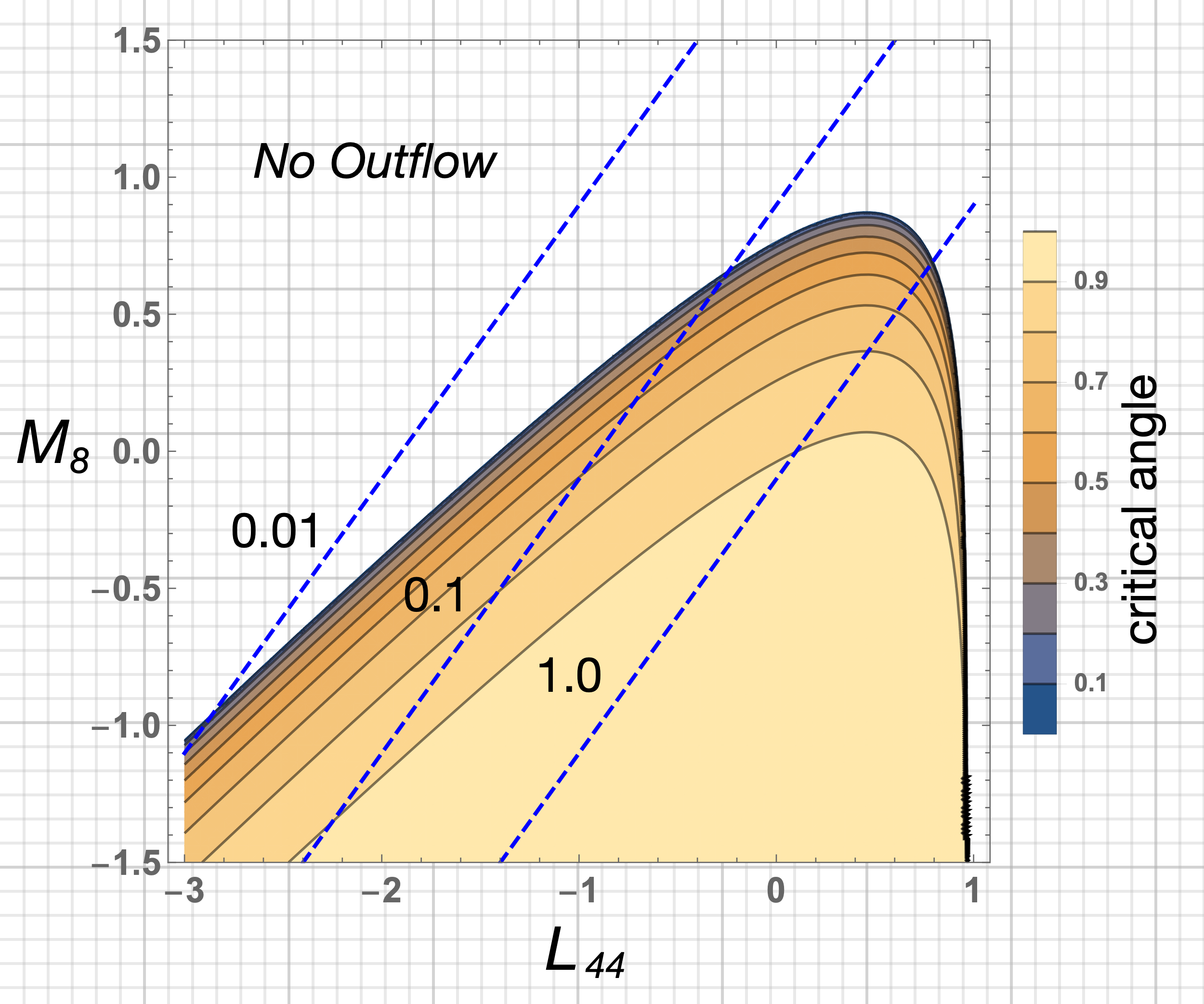} 
\caption{The critical angle of the radiation-accelerated gas $\theta_{crit}/(\pi/2)$ (eq. (\ref{eqn: 3-8})) as a function of 
the bolometric luminosity $L_{\rm 44} \equiv L_{\rm bol}/10^{44}\, {\rm erg}\,{\rm s}^{-1}$ and 
the black hole mass $M_{8}\equiv M_{BH}/10^{8}M_{\odot}$.
In the colored region, $r_{max}(\theta)$ has a finite value (e.g. models 4 and 5 in Fig. 5), 
bipolar outflows are expected, 
and $\theta_{crit}$ roughly corresponds the half-opening angle of the outflows.
Therefore for a larger $\theta_{crit}$, $f_{obs}$ tends to be smaller.
The three dashed lines indicate the Eddington ratios ($\gamma_{Edd}$) of 1.0, 0.1, and 0.01.
Note that for a given $\gamma_{Edd}$, $\theta_{crit}$ is not constant, suggesting
the obscuring nature is not determined only by $\gamma_{Edd}$ (see discussion in \S 4).}
\label{wada_fig: theta_c}
\end{figure}

Let us first assume that the X-ray heating causes 
the inner part of the gas disk
to be a quasi-stable, uniform sphere with 
radius $r_{0}$, as schematically shown in Fig. 5.
In the gas sphere with $r=r_{0}$, the radiative cooling with cooling rate $\Lambda_{cool}$
is assumed to be balanced with the X-ray heating rate:
$
n_{g}^{2} \Lambda_{cool}  \sim  L_{\rm X}/(4\pi r_{0}^{3}/3),
$
where $n_{g}$ is the average gas number density. Then, the radius $r_{0}$ within which 
the X-ray heating is effective is
\begin{eqnarray}
r_{0} \sim (3 L_{\rm X}/4 \pi n_{g}^2 \Lambda_{cool})^{1/3}.  \label{eqn: 3-2}
\end{eqnarray}

Now, let us assume that the gas at $r > r_{d}$ is accelerated due to the radiation pressure of
the dust, where $r_{d}$ is the dust sublimation radius\footnote{
In the following discussion, we assume that $r_{0} > r_{d}$, which 
is always the case for $L_{AGN} < 10^{47}$ erg s$^{-3}$ with the parameter 
chosen here.}. Therefore, 
for the velocity $v_{0}$ at $r=r_{0}$, energy conservation requires
\begin{eqnarray}
\frac{1}{2}\left(v_{0}^2 - v_{max}^{2}\right)=  G M_{\rm BH} \left(\frac{1}{r_{0}}  - \frac{1}{r_{max}} \right),
\label{eqn: 3-3}
\end{eqnarray}
where $r_{max}$ is the maximum radius that the launched gas 
reaches at maximum velocity $v_{max}$.
For the dusty gas with opacity $\kappa$ and dust-to-gas ratio $\gamma_{d}$,  
the gas can be accelerated toward $v_{max}$ due to radiation pressure from $r_{0}$ to $r_{max}$: 
\begin{eqnarray}
v_{max}^{2} \sim   \frac{\kappa \gamma_{d} L_{AGN}(\theta)}{4\pi c } \left(\frac{1}{r_{0}} - \frac{1}{r_{max}} \right),  
\label{eqn: 3-4}
\end{eqnarray}
where $L_{AGN}(\theta)$ is the non-spherical radiation field of the AGN
depending on the angle from the rotational axis $\theta$. 
Similarly, $v_{0}$ is the velocity attained by to the radiation pressure from $r_{d}$ to $r_{0}$, i.e.,
\begin{eqnarray}
v_{0}^{2} \sim   \frac{\kappa \gamma_{d} L_{AGN}(\theta)}{4\pi c } \left(\frac{1}{r_{d}} - \frac{1}{r_{0}} \right).
\label{eqn: 3-5}
\end{eqnarray}
Using eq. (\ref{eqn: 3-3}), (\ref{eqn: 3-4}), and (\ref{eqn: 3-5}), we obtain $r_{max}$ as
a function of $L_{AGN}$ and $M_{BH}$:
\begin{eqnarray}
r_{max}(L_{AGN}, M_{BH}) &=&   \;\;\;\;\;\;\;\;\;\;\;\; \nonumber \\
 \frac{\hat{L}_{AGN} + GM_{BH}}{GM_{BH}/r_{0} - \hat{L}_{AGN} (1/r_{d} - 2/r_{0})},
\label{eqn: 3-6}
\end{eqnarray}
where $\hat{L}_{AGN} \equiv \kappa \gamma_{d} L_{AGN}/(8\pi c)$.
If $r_{max} \rightarrow \infty$, the velocity of the dusty gas becomes larger than the escape velocity, and
 we expect that an outflow is formed. 

For simplicity, we take $L_{AGN}(\theta) =  1/2 L_{UV}|\cos(\theta)|$, rather than 
$ L_{AGN}(\theta) \propto \cos \theta (1 +
2 \cos \theta)$ as in the numerical simulations, where $L_{UV}$ is 
the UV luminosity of the AGN. This
 difference is not essential for the following discussion.
If $ GM_{BH}/r_{0} = \hat{L}_{AGN} (1/r_{d} - 2/r_{0})$ in
 eq. (\ref{eqn: 3-6}), then $r_{max}$ tends toward infinity
for the critical angle $\theta_{crit}$, which is derived as
\begin{eqnarray}
\cos \theta_{crit} &=& \frac{GM_{BH}}{r_{0}} \frac{16\pi c}{\kappa \gamma_{d} L_{UV}} \left( \frac{1}{r_{d}} - \frac{2}{r_{0}}  \right)^{-1}\\
&=& \frac{\sigma_{T}}{m_{p}\kappa \gamma_{d} \, \gamma_{Edd} \, r_{0}} 
\left( \frac{1}{r_{d}} - \frac{2}{r_{0}}  \right)^{-1},
\label{eqn: 3-8}
\end{eqnarray}
where $\gamma_{Edd}$ is the Eddington ratio, $\sigma_{T}$ is the Thomson  cross-section, 
and $m_{p}$ is the proton mass. We expect that, for $\theta < \theta_{crit}$, the radiation force is sufficiently
large to 
allow the dusty gas to escape.
Thus, we expect
that, for this cone region, outflows of gas are formed. Otherwise,
the gas launched from the inner region can only reach $r =r_{max}(\theta)$.
We suppose that this gas eventually falls back toward the equatorial plane, 
which corresponds to the backflows seen in the hydrodynamic simulations (Fig. 1a and \citet{wada2012}).

Since both the equilibrium radius $r_{0}$ (eq. (\ref{eqn: 3-2})) and 
the dust sublimation radius $r_{d}$ depend on the luminosity of the AGN (e.g., $r_{d} \propto L_{AGN}^{0.5}$), 
 $r_{max}$ and $\theta_{crit}$ cannot simply be functions of the Eddington
ratio $\gamma_{Edd}$, but they depend 
on both the AGN luminosity and the black hole mass.
For non-spherical radiation, $r_{max}$ is also a function of the angle $\theta$. 
In Fig. 6, we plot $r_{max}(\theta)$ for five different parameter sets of $L_{46}\equiv 10^{46}$ ergs s$^{-1}$ and $M_{BH,8} \equiv 10^{8} M_{\odot}$, showing the maximum radius of the launched gas.
Here, we assume that $r_{d} = 1.3 L_{46}^{1/2} $ pc \citep[e.g.,][]{lawrence1991} and 
$\Lambda_{cool} = 10^{-22} \; {\rm erg}\; {\rm cm}^{3} \,{\rm s}^{-1}$.
For ($L_{46}, M_{BH,8}) = (0.02, 0.1)$ and $(0.2,1.0)$ (red and blue lines),  $r_{max}(\theta)$ has finite values
for some angles; thus,
conical outflows are expected for $\theta < \theta_{crit}$ .
In the remaining three models with $(L_{46}, M_{BH,8}) = (2.0, 10.0), (10.0, 10.0)$,
and $(0.01, 1.0)$, the accelerated gas cannot escape at any angle.
In these cases, we suspect that steady outflows are not formed or they are stalled.

In Fig. 7, we plot $\theta_{crit}$ (eq. (\ref{eqn: 3-8})) as a function of $M_{BH}$ and $L_{bol,44}$.
Constant Eddington ratios are also shown with dotted lines. 
 Here, we assume $\kappa  = 10^{3} \,{\rm cm}^{2}{\rm g}^{-1}$ and
$\rho_{g} = 300 M_{\odot}\; {\rm pc}^{-3}$. 
Figure 7 shows that there are regions where no outflows are formed, i.e., 
$\theta_{crit}$ has no value.
For a given black hole (BH) mass ($M_{BH} \lesssim 10^{9} M_{\odot}$), 
there is a lower limit of the luminosity forming the outflows (i.e., $\pi/2 > \theta_{crit} > 0$),
under which the radiation pressure for the dusty gas is not sufficiently large to make
the outflow velocity larger than the escape velocity. 
Interestingly, there is also an upper limit of the luminosity forming the outflows.
This limit occurs because the dust sublimation 
radius $r_{d}$ becomes close to the cooling radius $r_{0}$; as a result,
the factor $(1/r_{d} - 2/r_{0})^{-1}$ is dominant in eq. (\ref{eqn: 3-8}). 

As we found in the radiation-driven fountain in the radiation-hydrodynamic simulations
(\citet{wada2012} and \S 2, part of the bipolar outflows driven by radiation from the AGNs
falls back to the disk plane, forming a thick disk (the torus).
The opening angle of the outflow can then be simply assumed to be close to the
critical angle $\theta_{crit}$ derived in the above argument. 
Thus, for larger $\theta_{crit}$, the obscuration fraction $f_{obs}$ is expected to be smaller. 
On the other hand, if $r_{max}$ is not infinity for all of the angles, 
the outflow is not formed. In this case, we expect that the central source is visible for
most viewing angles\footnote{The hydrodynamic simulations actually show more complicate 
dynamics of the gas for marginal cases, that is the gas shells are launched and falling back showing intermittent
behavior.}. 

Based on these simple arguments, we can qualitatively
understand the numerical results in \S 3.
We found that the obscuration fraction for a given column density, 
especially for $N_{22}$, increases 
with the luminosity of the AGN, and turns to be a small value 
above $L_{\rm X} \gtrsim 10^{45} $ erg s$^{-1}$.
The increase of obscuration is consistent with the behavior of the opening angle
as a function of the central luminosity (Fig. \ref{wada_fig: theta_c}). 
If the BH is too massive, e.g., $M_{BH}> 10^{9}M_{\odot}$, however
the outflow is not expected even for large luminosity. 
This situation occurs in the numerical simulations, and 
as a result, we observe a relatively small obscuration fraction for
high luminosity AGNs.

In the simplest version of the receding torus model \citep{lawrence1991, hasinger2008}, 
a constant scale height is assumed for the gas disk. Then, the dust sublimation radius, which
is a function of the central luminosity, is the only factor affecting the
opening angle. However, our radiation hydrodynamics calculations and analytic argument
provide a more dynamic picture. The radiation from the AGNs naturally drives
outflows if conditions are satisfied, which affect the obscuring nature. The response of the dust to the 
central radiation 
is important, but this response would be more dynamically
related with the obscuration than that assumed in the static receding torus model.

Note, however, that strong outflows are observed in some
luminous quasars, which could originate in the line-driven wind from 
the accretion disk \citep{elvis2000,proga2000, nomura2013}.
The line-driven outflows also contribute to the obscuration and could be
related with ultra-fast outflows (UFOs).

\begin{figure}[h]
\centering
\includegraphics[width = 6.5cm]{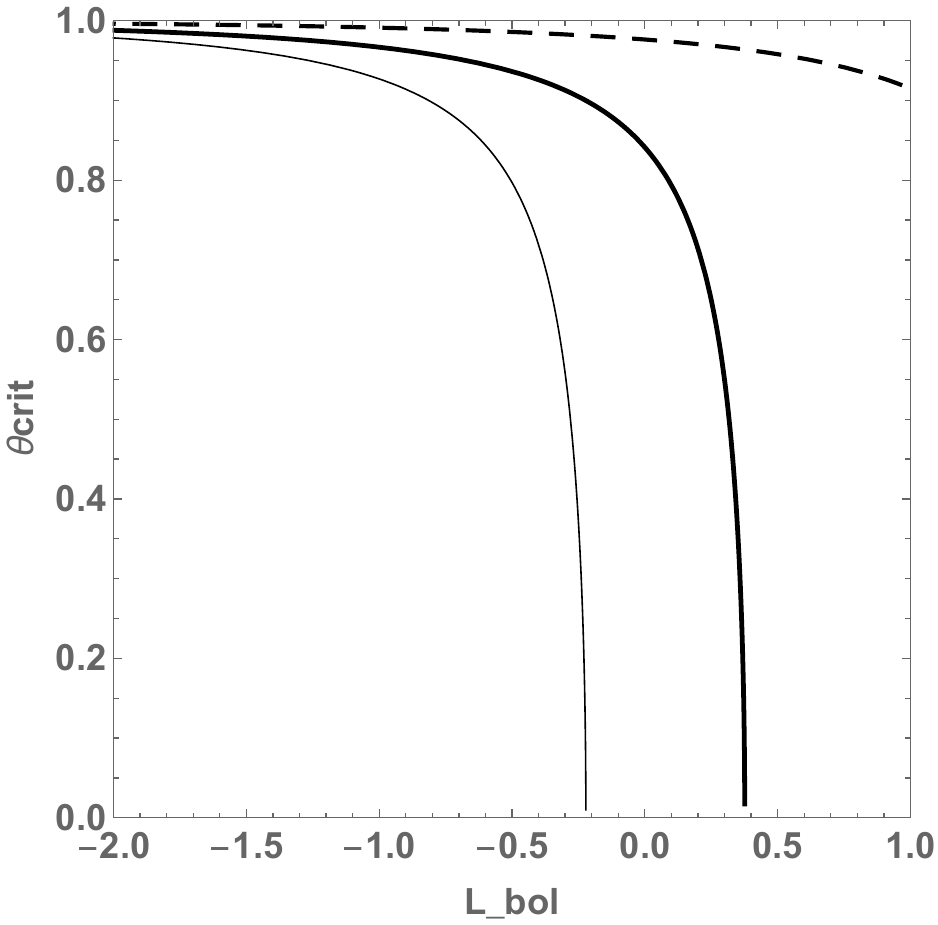} 
\caption{Opening angle of the outflows as a function of the bolometric luminosity for
three different gas densities $\rho_{g}(\times 10^{2}\, M_{\odot}\, {\rm pc}^{-3})$ = 1.0, 5.0, and 10.0  (dashed line, thick solid line, and thin solid lines, respectively), 
assuming a constant Eddington ratio of $0.2$.}
\label{wada_fig: f009}
\end{figure}

\section{IMPLICATIONS TO OBSERVATIONS}
\subsection{Torus structures inferred from $f_{obs}$}
In the simplest picture of the torus paradigm, the tori for all AGNs
have the same geometry, and the observed fraction of type-1 and 2 AGNs 
can be used to infer the opening angle of the torus. However, there is increasing evidence that 
the obscuring material around the nucleus depends on the properties of the AGNs.

The covering fraction of AGNs due to 
the surrounding gas can be estimated
by  the fraction of the reprocessed infrared radiation to
the bolometric luminosity of AGN \citep{maiolino2007, treister2008, lusso2013}, 
or from the fraction of type-2 AGNs based on
X-ray spectroscopy \citep{ueda2003,hasinger2008}.
However, there is discrepancy in the fraction among the observational results. The obscuring fraction inferred from optical and infrared observations is relatively high ($f_{obs} \sim 0.4-0.6$) \citep{lawrence2010,roseboom2013} and shows weak dependence on the luminosity of AGNs, 
whereas X-ray samples suggest that $f_{obs}$ decreases (from $\sim 0.8$ to $\sim 0.1$) with 
increasing luminosity in the $L_{\rm X} = 10^{42-46}$ erg s$^{-1}$ range \citep{lawrence2013}.
Based on 513 type-1 AGNs in the $XMM$-COSMOS survey, \citet{lusso2013}  show that 
the obscuring fraction ranges between $f_{obs} = 0.75-0.45$ or $0.45-0.30$, depending on 
the assumed physical properties of the torus, especially on its optical thickness\footnote{Note that the SED-based obscuring fraction does not necessarily 
represent the true value.
For example,  \citet{schartmann2014} derived
$f_{obs}$ directly from 
the optical depth calculated from the numerical results of 
the radiation-driven fountain model \citep{wada2012}.
They compare their results with $f_{obs}$ derived using a similar method by \citet{lusso2013} (see \citep{lusso2013}
and their Fig. 12) and found that the SED-based $f_{obs}$ (0.7) is effectively smaller than the 
directly determined value (0.9).}.  
The optically thick case seems consistent with the receding torus model \citep{simpson2005},
but the scale height of the torus is required to weakly increase as a function of the bolometric luminosity to fit the observed results.
Based on {\em WISE-UKIDSS-SDSS} data, \citet{roseboom2013} found that 
the ratio $L_{IR}$/$L_{bol}$, which should be related to the covering factor (CF),
shows a log-normal distribution around CF $\sim 0.3-0.4$ for luminous quasars.
This might represent the variety in the obscuring material structures.

Recently, based on the analysis of AGNs at $z < 0.2$ from {\em SDSS DR7}, \citet{oh2015} suggested
that 
the type-1 AGN fraction depends on both the black hole mass and bolometric luminosity,
shaping a ``ridge'' where the fraction of type-1 AGNs is low ($< 0.2$) 
on the plane of $L_{bol}-M_{BH}$ (see Fig. 16 in \citet{oh2015} ). 
%
For a given $M_{BH}$, the obscured fraction is larger for
less luminous, sub-Eddington AGNs. These results are roughly consistent with 
the analysis in \S 4 and Fig. \ref{wada_fig: theta_c}.

These observations and our results presented 
here suggest that the obscuration of the AGNs
should be a much more complicated phenomenon than the standard picture based on 
a static, dusty torus or a disk \citep[see also a recent review by][]{netzer2015}.
The discrepancy between X-rays and longer wavelength radiation in terms of
$f_{obs}$ implies that multiple components of the ISM contribute
the obscuration on various scales. Our numerical simulations showed that
radiative feedback causes non-steady, non-uniform structures of the 
multi-phase gas around the nucleus, and 
these structures are affected by at least the luminosity of AGNs and the black hole mass.
We should note, however that this situation could be even more complicated because many AGNs 
are associated with nuclear starbursts \citep[e.g.][]{esquej2014}, which
could also contribute to the obscuration \citep{wada2002}.

\subsection{Origin of luminosity-dependent obscuration in X-rays}
We now focus on the X-ray properties of the obscuration.
Recent X-ray spectral surveys based on 
large samples have reported that the fraction of the obscured objects changes
depending on the AGN luminosity. In other words, $f_{obs}$ rapidly decreases 
as the luminosity increases \citep{hasinger2008, ueda2003, burlon2011, brightman2011,ueda2014, merloni2013,aird2015}; 
the absorbed AGNs are less than $\sim$  30\%  for luminous AGNs at $L_{\rm X} \sim 10^{45}$ erg s$^{-1}$,
whereas the fraction of the absorbed AGNs is more than 70\% for less luminous ones at $\sim 10^{43}$ erg s$^{-1}$.
Several studies have also claimed that the fraction of obscured AGNs 
{\it increases}  with $L_{\rm X}$ \citep{burlon2011, brightman2011, buchner2015} for
lower luminosities. These results suggest that the fraction of the obscured AGNs in low redshift objects
peaks ($f_{obs} \sim 0.70-0.8$) 
around $L_{\rm X} \sim 10^{42.5-43}$  erg s$^{-1}$.

 The behavior of $f_{obs}$ in terms of the X-ray luminosity
  is basically consistent with the results shown in \S 3, even though 
  we found that AGNs with $L_{\rm X} \sim 10^{43-44}$ erg s$^{-1}$ are 
 the most effectively obscured, which is one order of magnitude larger than
 those suggested by observations.
Interestingly, this behavior rather fits high redshift ($z > 1.5-2.1$ in \citet{buchner2015}) 
  objects (see \S 5.3). 
   

\citet{buchner2015} schematically illustrated how the obscured fraction depends on 
the luminosity (see Fig. 13 in \citet{buchner2015} ). In particular, they suggested that at 
low luminosities, {\it torus clouds
are produced by the disk wind} \citep[e.g.,][]{elitzur-shlos2006}, and
{\it a receding torus caused by cloud ionization} \citep[e.g.,][]{lawrence1991} is 
proposed for the reason of the declining obscuring fraction for luminous objects.
These pictures are partly similar to results we have shown here, but the outflows are not
actually caused by the disk wind, which is a much smaller scale 
phenomenon (see, for examplem \citet{proga2000, nomura2013}) than that considered here.
In our radiation-hydrodynamic models, dusty gas irradiated by the central radiation source 
and its attenuation are implemented in the circumnuclear region.
However, in contrast to the original receding torus model, 
the response of the dusty gas is much more dynamic. 
\citet{davies2015} analyzed 20 AGNs observed by {\it Swift/BAT} as well as 
results by \citet{merloni2014}, which is based on 1310 AGNs observed 
{\em XMM-COSMOS} survey. They found that optical and X-ray properties of the absorption are different
depending on the luminosity. They proposed a modified version 
of the receding torus with a {\em neutral} torus inside the dusty molecular torus
by which the X-rays are absorbed.
In our simulations, the strong radiation from the accretion disk 
causes the dust-free region inside the dense dusty gas, and 
this also affects generation of the outflows.

\citet{buchner2015} also suggested that Compton-thick AGNs occupy about half of the 
the number and luminosity densities of the obscured population. The fraction of Compton-thick AGNs to
the total population is constant ($\sim $ 35\%) and is independent of the redshift and accretion luminosity.
In our numerical models, $f_{obs}$ for a larger column density (i.e., $N_{23}$) slightly increases from $\sim$ 15\% to 20\% over
the luminosity range of $L_{\rm X} = 10^{42-45}$ erg s$^{-1}$, and there is no
peak around $L_{\rm X} \sim  10^{44}$ erg s$^{-1}$, which contrasts the behavior of $f_{obs}$ for $N_{22}$. 
This occurs because the obscuring structures around AGNs basically consist of 
three components: a geometrically thin cold disk, a thick warm disk, and non-steady bipolar outflows \citep{schartmann2014}. The Compton-thick obscuration is mainly caused by dense, thin disks.
As a natural consequence, there are fewer chances to observe nuclei with higher column densities
(see Fig. 11 and 12 in \citet{buchner2015}).
\citet{jia2013} claimed that the column density obtained from X-ray absorption
 is nearly constant ($\sim 10^{22.9}$ cm$^{-2}$) in 71 type-2 quasars. 
 Our results suggest that $f_{obs}(N_{23})$ is 20--30\% 
 for the most luminous models with $L_{bol} > 10^{46}$ erg s$^{-1}$.
 The circumnuclear gas disks in these models are similar without showing strong outflows, suggesting that we have fewer chances to observe 
 luminous quasars with a {\em mild} obscuration with small column density (e.g., $< 10^{22}$ cm$^{-2}$). 

%
%


\subsection{Implication to the origin of the redshift evolution}

The X-ray surveys discussed in the previous section also showed that obscured AGNs (Compton-thin AGNs) increase toward $z=3-5$.
For example, \citet{buchner2015} showed that the fraction of the absorbed AGNs is $\sim 0.7$  and $\sim 0.6$ at $L_{X,44} \simeq 1$ and 10, respectively, at z=3 in contrast to $\sim 0.4-0.5$ at $z=0.5$.
 \citet{aird2015} also suggested similar results and claimed that $f_{obs}(L_{\rm X})$ differs depending on 
the luminosity at high-z. These studies also suggested that
the fraction of Compton-thick objects is nearly constant ($\sim$ 0.4 or 0.2) toward $z=3-5$.

Figure \ref{wada_fig: f009} shows the opening angle of the bipolar outflows
derived from the same analytical model as that shown in Fig. \ref{wada_fig: theta_c}.
In this plot, $\theta_{crit}$ is plotted for three initial densities of the gas disk, i.e., $\rho_{3} = 1.0, 5.0,$ and 10,
commonly showing that the opening angle becomes small as the central luminosity increases.
Moreover, Fig. \ref{wada_fig: f009} shows that, for a given luminosity, $\theta_{crit}$ is smaller for denser gas, i.e., more obscured.
If the gas density in the central region is not {\it too dense} (otherwise, 
there is no solution for $\theta_{crit}$, meaning that outflows are not formed),
AGNs could be {\it more obscured for more gas-rich environments}.
If high-z AGNs are associated with denser gas compared to  local objects,
this result  is qualitatively consistent with the trend shown by the X-ray survey.
Note again that this results is suggested only by the framework of the radiative feedback.
In reality, we suspect that, in very gas-rich objects at high-z, 
nuclear starbursts should be initiated, which 
could be also play a role in the obscuration through
supernova feedback \citep[e.g.][]{levenson01,wada2002}.


%
\section{SUMMARY AND REMARKS}
Based on the radiation-driven fountain scenario \citep{wada2012, schartmann2014}, which
is the circulation of gas driven by the central radiation obscuring 
the nucleus, we investigated how the AGN properties, i.e., luminosity and
black hole mass, affect the obscuration due to the circumnuclear ISM,  
using fully three-dimensional radiation-hydrodynamic simulations.
The obscuring fraction $f_{obs}$ is
derived from distribution of column density $N_{\rm H}$ toward the nucleus, 
using the numerical simulations. 

We found that $f_{obs}$ changes
depending on $N_{\rm H}$, the total luminosity of the
AGN ($L_{AGN}$), and the black hole mass ($M_{BH}$) (\S 3 and Fig. 6).
For a given $L_{AGN}$, $f_{obs}$ is always
smaller ($\sim 0.2-0.3$) for large column density ($N_{\rm H} > 10^{23}$ 
cm$^{-2}$) than that for a smaller column density $N_{\rm H} > 10^{22}$ cm$^{-2}$.
This is a natural consequence of the fact that the dense gas surrounding the nuclei is mostly
concentrated on a thin disk. 

We found that  $f_{obs}$ for
$N_{\rm H} = 10^{22}$ cm$^{-2}$ increases from $\sim 0.2$ to $\sim 0.6$ 
with the X-ray luminosity $L_{\rm X}$ in a range of $L_{\rm X} = 10^{42-44}$ ergs
s$^{-1}$, and  $f_{obs}$ becomes small $\sim 0.4$ at high luminosity
($\geq 2\times 10^{44}$ ergs s$^{-1}$).  
At a moderate luminosity $L_{\rm X} \sim 10^{44}$ ergs s$^{-1}$, 
a large fraction of the solid angles ($f_{obs} > 0.5$) can be obscured 
by the column density of $N_{\rm H} \leq 10^{22}$ cm$^{-2}$, because
of the outflows and thick disk caused by the back flows.
The $f_{obs}$ behavior has been simply demonstrated using a kinematic model (\S 4),
in which the bipolar outflows are formed in specific ranges of 
$L_{AGN}$ and $M_{BH}$ (Fig.5 and Fig. 6), and their opening angles do not
simply depend on the Eddington ratio.

The AGNs can be buried with $f_{obs}> 0.7$ for $N_{\rm H} \leq
10^{22}$ cm$^{-2}$,  
when the central accretion disk is {\it inclined} by 45$^{\circ}$ from 
the rotational axis of the circumnuclear gas disk (Fig. 1d and 2d, and open symbols in Fig. 3).
Even in this configuration,
the radiative feedback from the accretion disk 
cannot destroy a dense gas disk from a few to tens of parsecs.
 This causes non-spherical circulations of the gas (Fig. 1d), and 
as a result, the majority of the solid angles are obscured by the gas with $N_{\rm H} \le 10^{22}$ cm$^{-2}$.
The mass accretion can be possible through the dense gas disk, which is not completely destroyed, 
even in this case. 
This behavior could be a possible structure of
``buried'' AGNs. 

The behavior of $f_{obs}$ as a function of 
the X-ray luminosity is consistent with recent X-ray
surveys \citep{burlon2011, buchner2015, aird2015} (\S 5.2). However, the luminosity for which $f_{obs}$ reaches its
maximum is a factor of ten smaller (i.e., $\sim 10^{43}$ erg s$^{-1}$) than the observations for local AGNs,
and the luminosity would more accurately fit those suggested in AGNs at $z=2-3$ (\S 5.3).

Note that we did not consider the effect of stellar feedback 
around AGNs in the present simulations and analysis.
Many AGNs are associated with circumnuclear starbursts \citep[e.g.,][]{esquej2014, netzer2015}, 
and high-velocity outflows of molecular and ionized gas are often observed in quasars and ultra-luminous infrared galaxies (ULIRGs).
These activities could also change the structure of the
circumnuclear disk, and as a result the obscuration of AGNs can be affected.


%
\vspace{0.5cm}
%
The author is grateful to R. Davies, M. Schartmann, Y. Ueda and N. Kawakatsu
for many helpful comments and discussions. 
The author also appreciates many valuable comments and suggestions by the anonymous referee.
The numerical
computations presented in this paper were performed on a Cray XC30 
at the National Astronomical Observatory of Japan.



\newpage
%
%
\setcounter{section}{0}

\renewcommand{\thesection}{A-\arabic{section}}

 \section{Appendix} 
\subsection{Mechanism of the Outflows}

In this complementary section, we discuss dynamics and physical origin of the "radiation-driven fountain".
One may notice in some models that vertical ($z$-direction) motions
of the gas are driven (Fig. 1). 
Since the radiation pressure on the dusty gas due to the central radiation source is solved only for 
the radial direction by the ray-tracing method in this paper, this is mainly due to the effect of the gas pressure gradient caused by the radiative and mechanical heating and 
also by the density gradient.
In Fig. 9, we show snapshots of density, temperature ($T_g$), Mach number ($\cal{M}$) for vertical direction (i.e. $\cal{M}$ $\equiv v_z/c_s$, where $c_s$ is
the sound speed), and pressure distributions on the $x-z$ plane of model 8D50 (the inclination of the central accretion disk
is zero). 
It shows that velocity of the gas in the bipolar outflows along the rotational axis ($z-$ axis) is largest.
This is natural because the flux due to the central source is largest for the direction.  
However, the directions of the outflows are not perfectly 
radial, especially near the central ``core'' (torus-like structure seen in the density map inside $r\sim $ 5 pc) 
and the edges of the conical outflows.
As seen in the temperature map (Fig. 9b), a hot core with $T_g \sim 10^7$ K is formed, which is
directly heated by the X-ray from the central source. Its size roughly corresponds to the Compton radius $r_c \equiv GM_{BH} m_p/kT_g \simeq 4 \; {\rm pc} (T_g/10^7\, {\rm K})^{-1}(M_{BH}/1.3\times10^8 M_\odot)$.
In general, thermally driven wind is expected where $ r > r_c$ \citep[e.g.][]{begelman1983,woods1996}. 
This suggests that the gas motion just outside the hot core would be affected by the pressure gradient. 
However, as shown in Fig. 9c and 9d, the gases in the conical outflows are low temperature and highly
supersonic ($\cal{M}$  $>10-20$) at least for the vertical direction. This implies that the thermal pressure 
cannot keep driving the outflow motion, and the radiation pressure plays a key role as discussed in section 4. 
This is an essential difference between the outflows observed here and the thermally driven ``disk wind'' from the accretion disk,
which is a much smaller scale phenomenon and they are launched from the disk surface \citep{woods1996}. 
The temperature of the gas disk in the present models are too low to generate the disk winds.

It is also interesting that the ``back flows'' are generated outside the cone, whose direction is not necessarily radial.
For example, the accreting
 gases change their direction near the disk plane, which is due to the thermal pressure 
 as suggested by the small Mach number and also due to the relatively small radiation pressure near the disk plane.
 Self-gravity of the gas disk may also contribute to disturb the radial flows.

 \begin{figure*}[h]
\centering
\includegraphics[width = 12cm]{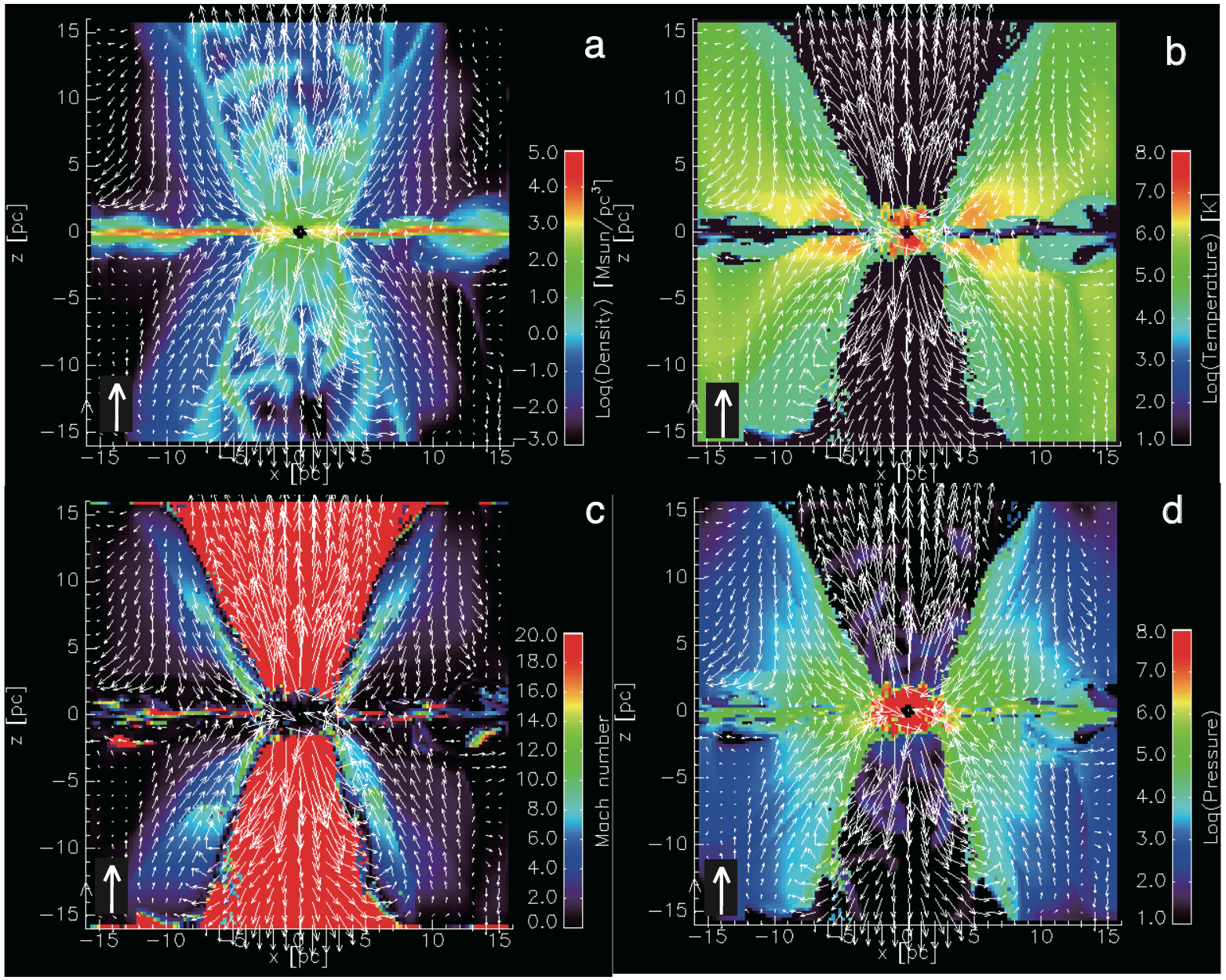} 
\caption{(a) density $\rho_g$, (b) temperature $T_g$, (c) Mach number for vertical direction, and (d) pressure $\equiv \rho_g T_g$
of model 8D50 at $t= 1.33$ Myr. Vectors are velocities on the x-z plane. A unit vector of 500 km s$^{-1}$ is shown 
in each panel. }
\label{wada_fig: A1}
\end{figure*}

\begin{figure}[h]
\centering
\includegraphics[width = 8cm]{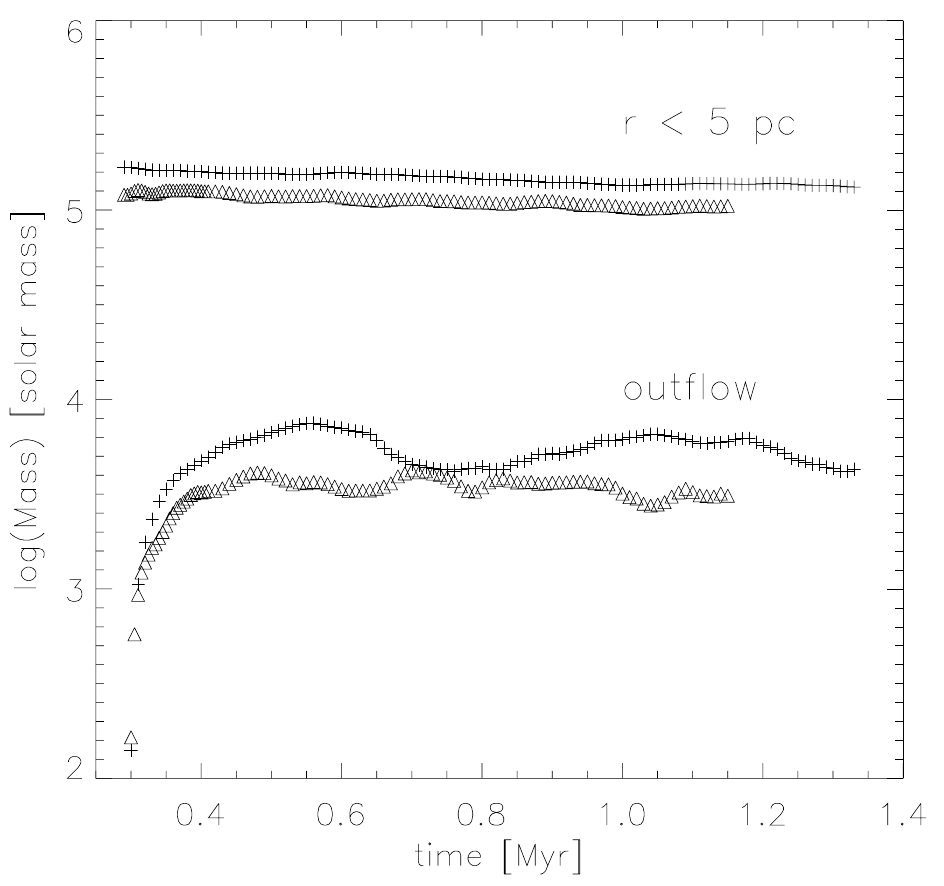} 
\caption{Evolution of mass of the outflows ($|z|> 5$ pc) and the core ($r < 5 $pc) in 
model 8D50 (triangles) and 8D30i (crosses).}
\label{wada_fig: mass_outflow}
\end{figure}

\subsection{Quasi-stability of the System and Mass accretion to the Central BH}

Although the outflows driven by the radiation pressure is locally non-uniform and non-steady, 
we assume in \S 3 and \S 5 that the obscured nature is based on globally quasi-steady 
structures of the circumnuclear gas over $\sim 1$ Myr. Figure 10 shows the gas mass 
in the outflows ($|z| >$ 5 pc) and the central core region ($r <$ 5 pc). 
It shows that the outflows are formed in the first $\sim$ 0.4 Myr, and a quasi-steady state is kept over
1 Myr. 
The gas mass in the central region is slightly decreasing with a rate of $\sim 0.2 M_\odot \; {\rm yr}^{-1}$, 
which is mostly caused by the outflows (see also discussion below).
The total gas mass in the system is about $6\times 10^6 M_\odot$, therefore the ``life time'' of the system would be
$\sim 3\times 10^7$ yr, if there is no mass supply from outside of the system.

Figure 11 shows time evolution of the gas mass just outside the inner boundary ($r \leq 0.75$ pc)
of model 8D50 and 8D30i.
It is significantly fluctuating (the time step is 50 years), and the distribution of the mass change rate $dM_g/dt$ is 
roughly Gaussian as shown in the histogram.  The root mean square (RMS) of $dM_g/dt$ is $\sim 0.3 M_\odot \; {\rm yr}^{-1}$.
Therefore if the short-period fluctuation of the gas mass caused by a temporal mass accretion toward the 
black hole and immediate by supply from the disk, the RMS accretion rate corresponds to about 10\% Eddington accretion rate,
which is comparable to the assumed luminosity of the central source.
However, the average accretion rate over a longer period is much smaller; 
 for example, for the ``accretion phase'' around $t=1$ Myr (Fig. 11), the accretion rate
is about $2\times 10^{-3} M_\odot \; {\rm yr}^{-1}$.
Therefore, the present simulations suggest that the radiation-driven fountain, whose time scale is at least $\sim$ Myr, 
cannot be sustained only by the mass accretion through the disk on a sub-pc scale.

The mass accretion rate could be enhanced if the angular momentum transfer in the disk
is more effective. This could be possible if the turbulent viscosity or 
gravitational torque due to the non-axisymmetric structures are enhanced due to
gravitational instability \citep[see for example][and references therein]{wada_meurer02}.
The Toomre's $Q$ value, $Q\equiv \kappa c_s/\pi G \Sigma_g$, 
where $\kappa$ is epicyclic frequency and $\Sigma_g$ is the surface density of the gas disk, is
%
\begin{eqnarray}
 Q &\simeq& 0.5 \left(\frac{c_s}{1\; {\rm km}\, {\rm s}^{-1}}\right) 
\left(\frac{\Sigma_g}{1000\; M_\odot \;{\rm pc}^{-2}} \right)^{-1} \nonumber \\
&\times& \left(\frac{M_{BH}}{1.3\times 10^7 \; M_\odot}\right)^{1/2} 
\left(\frac{r}{10\; \rm pc} \right)^{-3/2}.  \nonumber
\end{eqnarray}
Therefore the disk around $M_{BH} = 1.3 \times 10^7 M_\odot$
can be gravitationally unstable even for most regions, if the temperature of the
gas is smaller than 100 K. 
The gas temperature inside $r \lesssim 5$ pc is much larger than $10^4$ K due to
the X-ray heating, which is assumed to be spherical, 
thus disk becomes stable and geometrically thick (see Fig. 9a and 9b).
On the other hand, outer disk ($r> 10$ pc) is expected to be gravitationally unstable
especially for smaller BHs.
The most unstable wavelength in the differentially rotating disk, 
$\lambda_{\rm J}\equiv 2c_s^2/G\Sigma_g \simeq $ 1 pc $(T_g/100 K)(\Sigma_g/1000 M_\odot )^{-1}$, 
therefore the current spatial resolution, 0.25 pc, is merginal to resolve 
the gravitational instability of the coldest, thin gas disk (see also Fig.12).
However, the ``effective'' sound velocity including the velocity dispersion of the
self-gravitationally unstable disk can be much larger than 1 km s$^{-1}$
\citep[see for example,][]{wada01, wada_meurer02}, and also  
by perturbation due to the back flows.
As a result, the outer disk cannot be very thin (see Fig. 9) and 
the current spatial resolution is fine enough to resolve its vertical structures.

In order to clarify how the spatial resolution affects the mass accretion toward the center
through gravitational instability, 
we run two additional test calculations, as shown by Fig.12.
 Since the radiation pressure does almost no effect on the gas disk itself due to the angle dependence of the radiation field and to save computational time, here I follow initial evolution of the disk without the radiation feedback. I take a vertically smaller computational box, i.e. $32^2 \times$ 8 pc, which is solved using $128^2\times 32$ or $256^2 \times 32$ grid points. 
 Although the dense disk is thinner in the model with 0.125 pc resolution, the spiral-like features originated in self-gravitational instability are resolved in both models. Evolution of the mass in the central region ($r \leq 0.75$ pc) is also plotted for
 the two different resolution. 
 We found that the average mass accretion rate is somewhat larger in the model with smaller resolution (0.125 pc), but the difference is not significant.

If the dense gas disk continues to a larger radius (e.g. $r \sim 100$ pc) with
similar surface density, 
the disk should be highly unstable, therefore 
larger mass inflow to the central region due to the non-axysymmetric structures or gravity-driven turbulence 
could be possible.
This is an interesting issue to be explored using larger computational box and finer resolution 
in the near future,  as well as the effect of supernova feedback \citep[e.g.][]{wada02}
with the radiative feedback. 

\begin{figure}[h]
\centering
\includegraphics[width = 8cm]{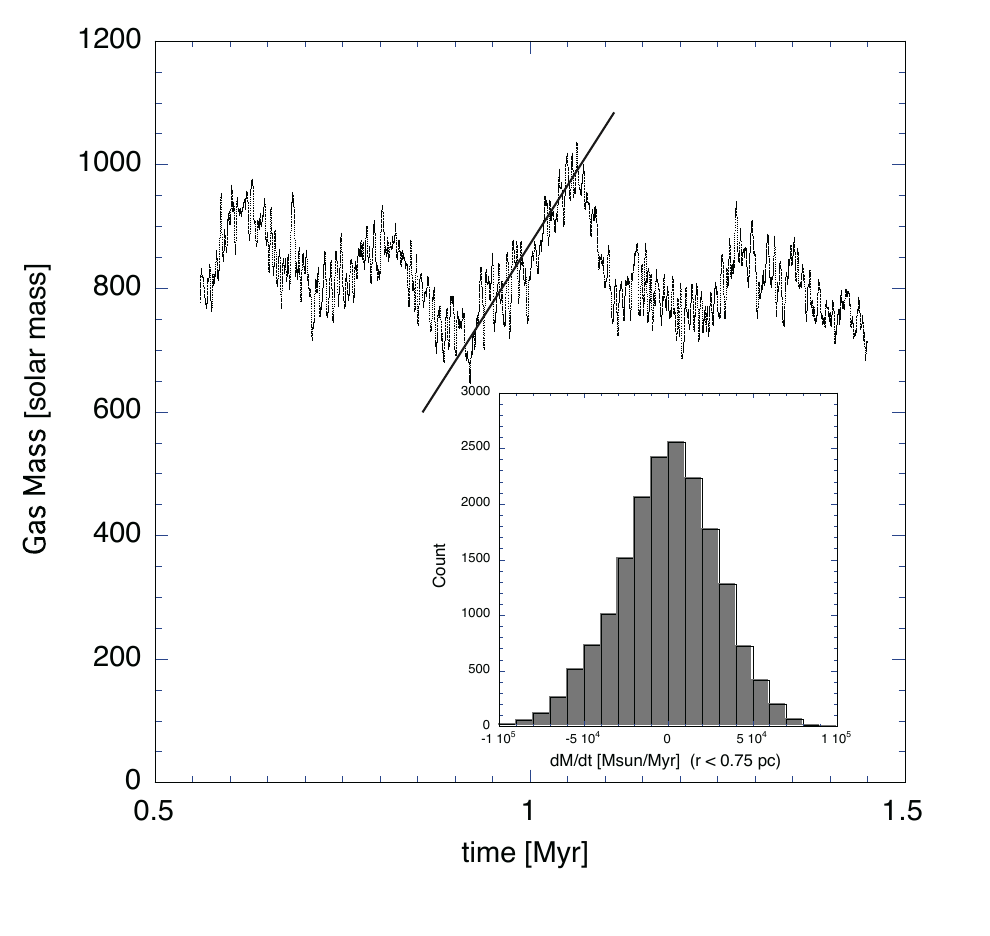} \\
\caption{Evolution of gas mass inside $r = 0.75$ pc and histogram of $dM/dt$ in model 8iD30i. The root mean square of $dM/dt$ is 
0.3 $M_\odot\; {\rm yr}^{-1}$.   The solid line around $t=1$ Myr
represents the average accretion rate of $2\times 10^{-3} M_\odot\; {\rm yr}^{-1}$.}
\label{wada_fig: mdot}
\end{figure}

\begin{figure*}[h]
\centering
\includegraphics[width = 6cm]{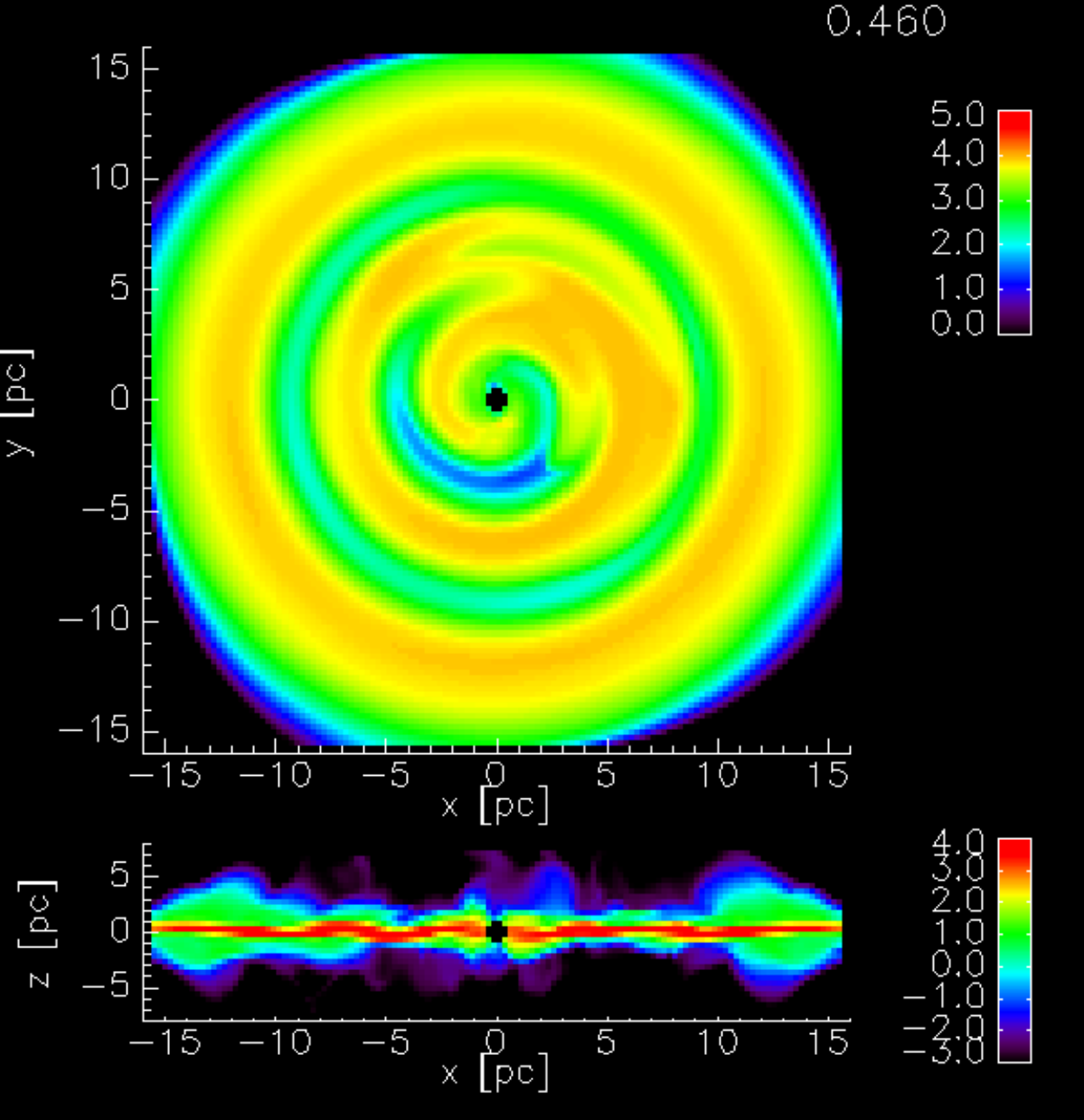}
\includegraphics[width = 6cm]{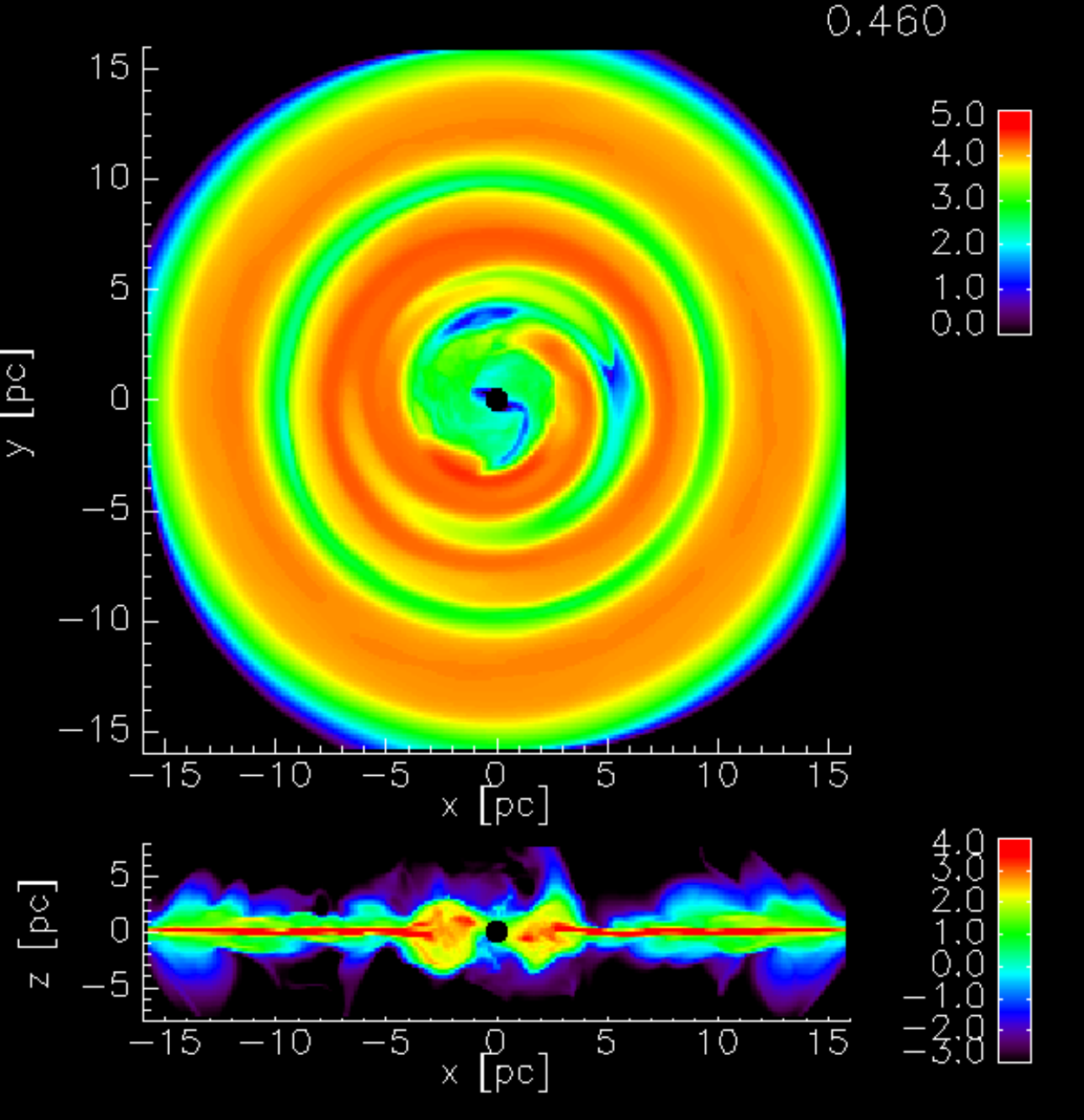} 
\includegraphics[width = 7cm]{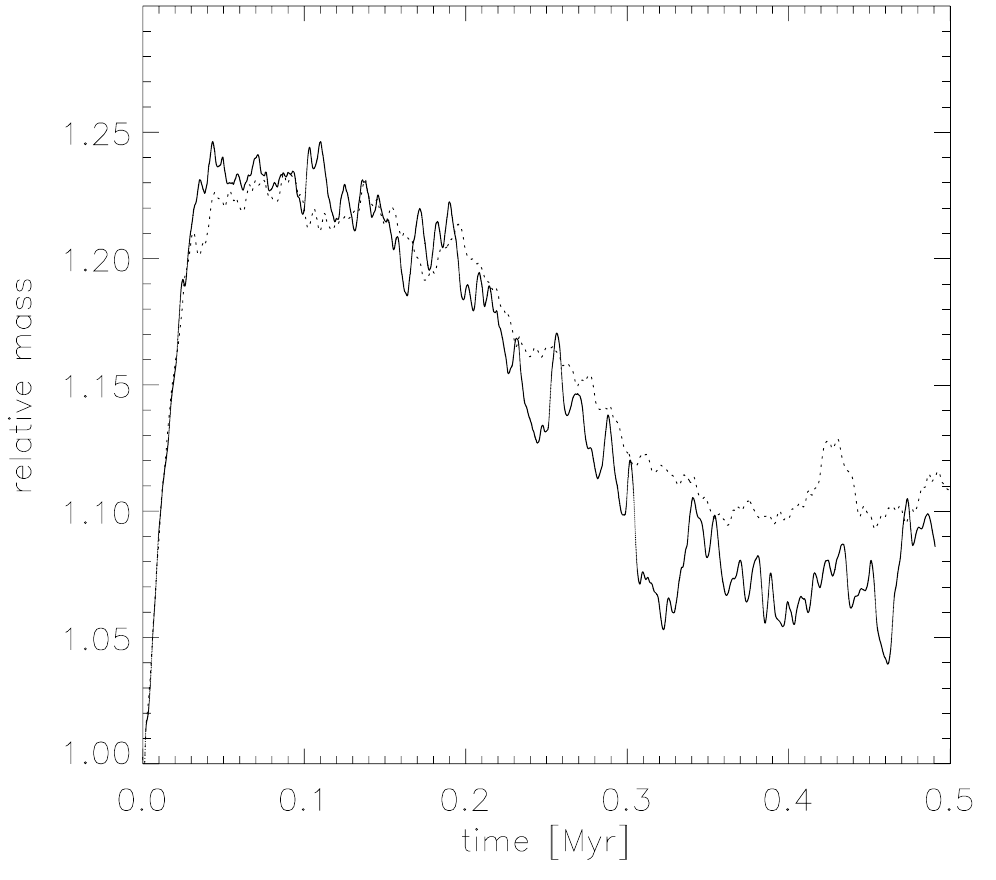} 
\caption{Test calculations of a disk without the radiation feedback in order to see effect of the
resolution to the mass accretion through the disk. The black hole mass is $1.3\times 10^7 M_\odot$ and the
initial density of the disk is $1.22\times 10^3 M_\odot \;{\rm pc}^{-3}$. (left) density for the model with 0.25 pc resolution at $t=0.46$ Myr (right) Same as the left panel, but with 0.125 pc resolution.
The color bar is log-scaled volume density ($M_\odot\; {\rm pc}^{-3}$).
(bottom) Evolution of the mass normalized to the initial value near the inner boundary ($r \leq 0.75$ pc). 
Thick and dotted lines are the runs with 0.125 pc and 0.25 pc resolution, respectively. }
\label{wada_fig: disk}
\end{figure*}

\newpage



\begin{thebibliography}{}


\bibitem[Aird et al.(2015)]{aird2015} Aird, J., Coil, A.~L., 
Georgakakis, A., et al.\ 2015, \mnras, 451, 1892 
\bibitem[Ballantyne et al.(2006)]{ballantyne2006} Ballantyne, D.~R., Shi, Y., Rieke, G.~H., et al.\ 2006, \apj, 653, 1070 

\bibitem[Begelman et al.(1983)]{begelman1983} Begelman, M.~C., 
McKee, C.~F., \& Shields, G.~A.\ 1983, \apj, 271, 70 
\bibitem[Brightman 
\& Nandra(2011)]{brightman2011} Brightman, M., \& Nandra, K.\ 2011, \mnras, 414, 3084 
\bibitem[Burlon et al.(2011)]{burlon2011} Burlon, D., Ajello, M., 
Greiner, J., et al.\ 2011, \apj, 728, 58 
\bibitem[Buchner et al.(2015)]{buchner2015} Buchner, J. Georgakakis, A., Nandra, K. et al.   ApJ 802, id89 (2015)
%
%
%
%
%
\bibitem[Davies et al.(2015)]{davies2015} Davies, R., Burtscher, 
L., Rosario, D., et al.\ 2015, arXiv:1505.00536 
%
\bibitem[Dorodnitsyn et al.(2012)]{dorodnitsyn2012} Dorodnitsyn, A., Kallman, T., \& Bisnovatyi-Kogan, G.~S.\ 2012, \apj, 747, 8 
\bibitem[Dorodnitsyn et al.(2011)]{dorodnitsyn2011} Dorodnitsyn, A., Bisnovatyi-Kogan, G.~S., \& Kallman, T.\ 2011, \apj, 741, 29 

%
\bibitem[Elitzur(2006)]{elitzur06} Elitzur, M.\ 2006, New Astronomy Review,		       50, 728
\bibitem[Elitzur \& Shlosman(2006)]{elitzur-shlos2006} Elitzur, M., \& Shlosman, I.\ 2006, \apjl, 648, L101 
\bibitem[Elvis(2000)]{elvis2000} Elvis, M.\ 2000, \apj, 545, 63 
\bibitem[Elvis(2012)]{elvis2012} Elvis, M.\ 2012, Journal of 
Physics Conference Series, 372, 012032   http://arxiv.org/abs/1201.5101
\bibitem[Enoki et al.(2003)]{enoki2003} Enoki, M., Nagashima, M., 
\& Gouda, N.\ 2003, \pasj, 55, 133 
\bibitem[Enoki et al.(2014)]{enoki2014} Enoki, M., Ishiyama, T., 
Kobayashi, M.~A.~R., \& Nagashima, M.\ 2014, \apj, 794, 69 
\bibitem[Esquej et al.(2014)]{esquej2014} Esquej, P., Alonso-Herrero, A., Gonz{\'a}lez-Mart{\'{\i}}n, O., et al.\ 2014, \apj, 780, 86 
\bibitem[Fanidakis et al.(2011)]{fanidakis2011} Fanidakis, N., Baugh, 
C.~M., Benson, A.~J., et al.\ 2011, \mnras, 410, 53 
\bibitem[Hasinger(2008)]{hasinger2008} Hasinger, G.\ 2008, \aap, 490, 905 
\bibitem[Han et al.(2012)]{han2012} Han, Y., Dai, B., Wang, B., Zhang, F., \& Han, Z.\ 2012, \mnras, 423, 464 

 
\bibitem[Hopkins et al.(2007)]{hopkins2007} Hopkins, P.~F., 
Richards, G.~T., \& Hernquist, L.\ 2007, \apj, 654, 731 
%

%
%
\bibitem[Jia et al.(2013)]{jia2013} Jia, J., Ptak, A., Heckman, T., \& Zakamska, N.~L.\ 2013, \apj, 777, 27 
\bibitem[Kawaguchi \& Mori(2010)]{kawaguchi2010} Kawaguchi, T., \& Mori, M.\ 2010, \apjl, 724, L183 

\bibitem[La Franca et al.(2005)]{lafranca2005} La Franca, F., Fiore, 
F., Comastri, A., et al.\ 2005, \apj, 635, 864 
\bibitem[Laor 
\& Draine(1993)]{laor1993} Laor, A., \& Draine, B.~T.\ 1993, \apj, 402, 441 
\bibitem[Lawrence 
\& Elvis(1982)]{lawrenceelvis1982} Lawrence, A., \& Elvis, M.\ 1982, \apj, 256, 410 
\bibitem[Lawrence(1991)]{lawrence1991} Lawrence, A.\ 1991, \mnras, 
252, 586 
\bibitem[Lawrence et al.(2013)]{lawrence2013} Lawrence, A., 
Roseboom, I., Mayo, J., et al.\ 2013, arXiv:1303.0219 
\bibitem[Lawrence \& Elvis(2010)]{lawrence2010} Lawrence, A., \& Elvis, M.\ 2010, \apj, 714, 561 

\bibitem[Levenson, Weaver, \& Heckman(2001)]{levenson01} Levenson, N.~A., Weaver, K.~A., \& Heckman, T.~M.\ 2001, \apj, 550, 230 
\bibitem[Liou \& Steffen(1993)]{liou93} Liou, M., Steffen, C., 1993,J. Comp. Phys., 107, 23
\bibitem[Liou (1996)]{liou1996} Liou, M., 1996, J. Comp. Phys., 129, 364

\bibitem[Lusso et al.(2013)]{lusso2013} Lusso, E., Hennawi, 
J.~F., Comastri, A., et al.\ 2013, \apj, 777, 86 
\bibitem[Marconi et al.(2004)]{marconi2004} Marconi, A., Risaliti, 
G., Gilli, R., et al.\ 2004, \mnras, 351, 169 
\bibitem[Maloney, Hollenbach, \& Tielens(1996)]{maloney96} Maloney, P.~R., 
Hollenbach, D.~J., \& Tielens, A.~G.~G.~M.\ 1996, \apj, 466, 561 
\bibitem[Maiolino et 
al.(2007)]{maiolino2007} Maiolino, R., Shemmer, O., Imanishi, M., et al.\ 2007, \aap, 468, 979 
\bibitem[Meijerink \& Spaans(2005)]{meijerink05} Meijerink, R., \& Spaans, M.\ 2005, \aap, 436, 397 
%
\bibitem[Merloni 
\& Heinz(2013)]{merloni2013} Merloni, A., \& Heinz, S.\ 2013, Planets, Stars and Stellar Systems.~Volume 6: Extragalactic Astronomy and Cosmology, p.503
\bibitem[Merloni et al.(2014)]{merloni2014} Merloni, A., Bongiorno, 
A., Brusa, M., et al.\ 2014, \mnras, 437, 3550 

\bibitem[Nenkova et al.(2008)]{nenkova2008} Nenkova, M., Sirocky, 
M.~M., Ivezi{\'c}, {\v Z}., \& Elitzur, M.\ 2008, \apj, 685, 147 

\bibitem[Netzer(1987)]{netzer1987} Netzer, H.\ 1987, \mnras, 225, 55 
\bibitem[Netzer(2015)]{netzer2015} Netzer, H.\ 2015, \araa, 53, 365 
\bibitem[Nomura et al.(2013)]{nomura2013} Nomura, M., Ohsuga, K., 
Wada, K., Susa, H., \& Misawa, T.\ 2013, \pasj, 65, 40 
\bibitem[Oh et al.(2015)]{oh2015} Oh, K., Yi, S.~K., Schawinski, K., et al.\ 2015, \apjs, 219, 1 
\bibitem[P{\^a}ris et al.(2014)]{paris2014} P{\^a}ris, I., Petitjean, P., Aubourg, {\'E}., et al.\ 2014, \aap, 563, A54 
%
%
\bibitem[Pier \& Krolik(1993)]{pier93} Pier, E.~A., \& Krolik, J.~H.\ 1993, \apj, 418, 673 
\bibitem[Proga et al.(2000)]{proga2000} Proga, D., Stone, J.~M., 
\& Kallman, T.~R.\ 2000, \apj, 543, 686 
\bibitem[Roseboom et al.(2013)]{roseboom2013} Roseboom, I.~G., 
Lawrence, A., Elvis, M., et al.\ 2013, \mnras, 429, 1494 
\bibitem[Shi \& Krolik(2008)]{shi-krolik2008} Shi, J., \& Krolik, J.~H.\ 2008, \apj, 679, 1018 
\bibitem[Schartmann et al.(2010)]{schartmann2010} Schartmann, M., 
Burkert, A., Krause, M., et al.\ 2010, \mnras, 403, 1801 
\bibitem[Schartmann et al.(2014)]{schartmann2014} Schartmann, M., 
Wada, K., Prieto, M.~A., Burkert, A., 
\& Tristram, K.~R.~W.\ 2014, \mnras, 445, 3878 
\bibitem[Shao et al.(2013)]{shao2013} Shao, L., Kauffmann, G., Li, C., Wang, J., \& Heckman, T.~M.\ 2013, \mnras, 436, 3451 
\bibitem[Simpson(2005)]{simpson2005} Simpson, C.\ 2005, \mnras, 
360, 565 
%
%
\bibitem[Treister et al.(2008)]{treister2008} Treister, E., Krolik, 
J.~H., \& Dullemond, C.\ 2008, \apj, 679, 140
\bibitem[Urry \& Padovani(1995)]{urry95} Urry, C.~M., \& Padovani, P.\ 1995, \pasp,			 107, 803 
\bibitem[Ueda et al.(2003)]{ueda2003} Ueda, Y., Akiyama, M., 
Ohta, K., \& Miyaji, T.\ 2003, \apj, 598, 886 
			 
\bibitem[Ueda et al.(2014)]{ueda2014} Ueda, Y., Akiyama, M., 
Hasinger, G., Miyaji, T., \& Watson, M.~G.\ 2014, \apj, 786, 104 		 
%
%
%
%
%
%
\bibitem[Wada(2001)]{wada01} Wada, K. 2001, \apjl, 559, L41
\bibitem[Wada(2012)]{wada2012} Wada, K. 2012, \apj, 758, 66 (Paper I)
\bibitem[Wada et al.(2002)]{wada2002} Wada, K., Meurer, G., \& Norman, C.~A.\ 2002, \apj, 577, 197 
\bibitem[Wada \& Norman(2002)]{wada02} Wada, K.~\& Norman, C.~A.\ 2002,				    \apjl, 566, L21 (WN02)
\bibitem[Wada et al.(2009)]{wada09} Wada, K., Papadopoulos, 
P.~P., \& Spaans, M.\ 2009, \apj, 702, 63 
\bibitem[Wada, Meurer, \& Norman(2002)]{wada_meurer02} Wada, K.,		       Meurer, G., \& Norman, C.~A.\ 2002, \apj, 577,		       197 
%
%
\bibitem[Woods et al.(1996)]{woods1996} Woods, D.~T., Klein, 
R.~I., Castor, J.~I., McKee, C.~F., \& Bell, J.~B.\ 1996, \apj, 461, 767 
\bibitem[Xu(2015)]{xu2015} Xu, Y.-D.\ 2015, \mnras, 449, 191 


\end{thebibliography}
\end{document}